**REVIEW**

# Practical Concepts for Design, Construction and Application of Halbach Magnets in Magnetic Resonance

**Peter Blümler[1]** · **Helmut Soltner[2]**

## Abstract

This review is a compilation of relevant concepts in designing Halbach multipoles for magnetic resonance applications. The main focus is on providing practical guidelines to plan, design and build such magnets. Therefore, analytical equations are presented for estimating the magnetic field from ideal to realistic systems. Various strategies of homogenizing magnetic fields are discussed together with concepts of opening such magnets without force or combining them for variable fields. Temperature compensation and other practical aspects are also reviewed. For magnetic resonance two polarities (di- and quadrupole) are of main interest, but higher polarities are also included.

## 1 Introduction

The invention of rare-earth magnets (typical properties are summarized in Appendix A) in the 1970s to 80s changed the way magnets could be designed, because they have much higher coercivities (magnetic "hardness") than formerly used AlNiCo or ferrite-permanent magnets. This fact allowed assembling magnetic flux sources similar to "toy blocks", because the magnets hardly impair their magnetic properties. Their high remanence and their low permeability permitted to design very strong but compact permanent magnets using analytical approaches. One of the pioneers in this field was John C. Mallinson, who together with Klaus Halbach invented what nowadays is referred to as "Halbach arrays". Their concept is based on a spatially

✉ Peter Blümler
  bluemler@uni-mainz.de

1   Institute of Physics, University of Mainz, 55099 Mainz, Germany

2   Zentralinstitut für Engineering, Elektronik und Analytik (ZEA-1), Forschungszentrum Jülich GmbH, 52425 Jülich, Germany

Springer

oscillating magnetization in rare-earth magnets that produces enhanced magnetic flux only on one of their sides. Generalization then led to cylinders, spheres etc. with extremely strong and homogeneous magnetic fields of variable polarity, which naturally found many applications in various fields.

The properties and uses of Halbach magnets have already been reviewed several times [1–3]. However, there has been only little focus on practical aspects in magnet construction. It is the aim of this work to show, how the magnetic fields of Halbach multipoles can be estimated and which design criteria are critical, how they can be built and what further options can be implemented. All this is done with emphasis on applications in magnetic resonance. During the last two decades Halbach dipoles have developed from tinkered lab-instruments to full commercial products[1] for benchtop NMR-relaxometry and spectroscopy. They work at magnetic fields from 1 to 3 T with homogeneities in the ppb-regime. Nevertheless, this field is still quite active in research, particular in developing portable MRI-scanners for the human head [4–7] for point-of-care application and to provide more affordable instrumentation for developing countries [8]. On the other hand, they have also been designed for EPR [9, 10], and DNP [10–12].

For magnetic resonance applications the interest in generating multipolar magnetic fields is typically limited to dipolar (for homogeneous magnetic fields) and quadrupolar (for homogeneous field gradients) arrangements. For completeness, however, higher polarities are touched at some points, but lengthy details are placed in appendices. The design of multipoles is interesting for particle traps in fundamental physics [13, 14], accelerators [15–18], motors [19], bearings [20], and linear machines [21]. Recently, they have also found use in techniques to guide magnetic nanoparticles remotely [2, 22] and in the emerging field of magnetic particle imaging, MPI [23].

This review is structured as follows: The concept of single-sided flux is introduced with Mallinson's planar array (Sect. 2), which is then intuitively generalized to cylindrical Halbach multipoles (Sect. 3). After their principal field is introduced, this ideal case is step by step applied to more realistic structures. Each of these steps is accompanied by analytical expressions for the consequences on the magnetic flux inside them. The discussion of possible demagnetization concludes this section, which allows a basic design and choice of suitable materials. Section 4 deals with various strategies to homogenize the field in the plane of the magnet and along its axis using passive, mechanical and active shim strategies. Finally, Sect. 5 introduces concepts to open Halbach multipoles without force, combine them for variable field sources, achieve temperature compensation and give some practical hints for manually assembling such magnets.

---

[1] NMR spectroscopy: e.g. Spinsolve 90 from magritek Ltd., Fourier 80 from Bruker Corp.

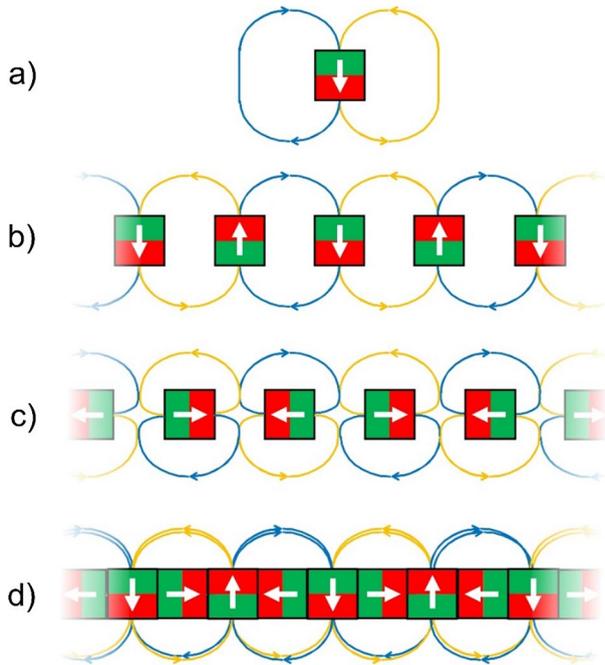

**Fig. 1** Schematic illustration of how magnets can be arranged to result in single-sided flux. **a** A single magnet with its magnetization direction downwards (indicated by the white arrow or the red and green color specifying north and south pole). The resulting dipolar magnetic field is represented by two field lines for the same field strength but different directions (blue = clockwise and gold = counterclockwise). **b** Several of the magnets from **a** arranged with a gap of the same size and alternating vertical magnetizations. **c** Same as **b** but now with alternating magnetization in horizontal direction. **d** Combination of the two arrangements in **b** and **c** gives a simplified planar Halbach array. It can clearly be seen how the flux cancels below the array and doubles on the other side (color figure online)

## 2 Planar Halbach Arrays

In 1973 Mallinson [24] published the idea of what he considered a magnetic curiosity then, a planar permanent magnet with a magnetization pattern that results in single-sided flux. The conceptual idea is illustrated in Fig. 1, in which magnets with quadratic cross sections are arranged such that their flux cancels on one side and increases on the other. An analytical expression for this one-dimensional magnetization pattern along $x$ is also given in the original publication

$$\boldsymbol{M}(x) = \begin{pmatrix} M_x \\ M_y \\ M_z \end{pmatrix} = M_0 \begin{pmatrix} \sin(k'x) \\ \cos(k'x) \\ 0 \end{pmatrix} \quad \text{with} \quad k' = \frac{2\pi}{\lambda}, \tag{1}$$

where $k'$ is the wave number (spatial frequency) or $\lambda$ the wavelength of the pattern (cf. Fig. 2a). The resulting field above the surface ($y > 0$) of such a structure of thickness $d$ (and infinite length in $z$) is then

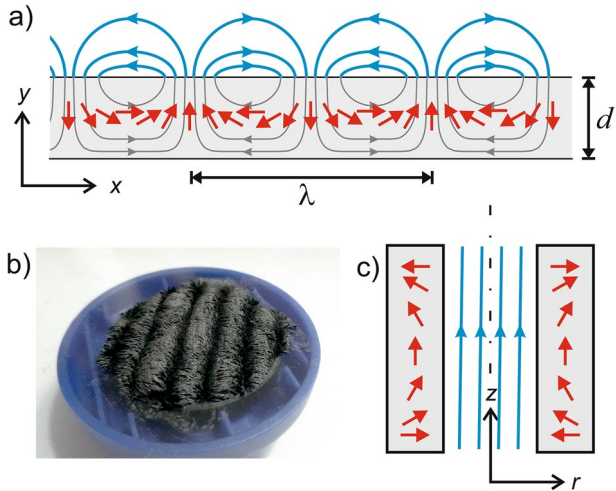

**Fig. 2** Linear, one-dimensional Halbach arrays: **a** a continuous version of the discrete scheme of Fig. 1d. The magnetic material (gray) has a magnetization (red arrows) pattern, which varies according to Eq. (1) along $x$ and is constant along $y$ over a thickness $d$, while it is infinite in the third dimension. This produces a recurring magnetic field (blue flux lines) on one side with wavelength $\lambda$. **b** Photograph of iron filings on a household magnet with ca. $3\lambda$ illustrating the flux pattern of **a**. The magnetic flux on the opposite side is negligible. **c** If a $\lambda/2$ section of such a magnet is rotated around an axis at some distance parallel to its surface, a hollow cylinder with axial field is produced. Under certain conditions this field can be quite homogeneous but is limited to certain aspect ratios. Note that this is NOT a typical Halbach cylinder as discussed in Sect. 3, since its field direction points in axial direction (color figure online)

$$\boldsymbol{B}(x,y) = \begin{pmatrix} B_x \\ B_y \end{pmatrix} = B_R \left(1 - e^{-k'd}\right) e^{-k'y} \begin{pmatrix} \sin(k'x) \\ \cos(k'x) \end{pmatrix}, \tag{2}$$

where $B_R$ is the remanence of the magnetic material (for typical values see Appendix A). The magnetic field components $B_x$ and $B_y$ are harmonic functions outside the magnet material. As a consequence, the sinusoidal variation in the horizontal direction goes together with an exponential decay in the vertical direction. Nowadays, many household magnets (see Fig. 2b) and magnetic foils have such magnetization patterns. This has the advantage that the holding force is roughly doubled per magnet weight and magnetic foils can be made thinner and more flexible. Other applications of such magnetization patterns are in the inductrack magnetic levitation [21] or wigglers and undulators to generate synchrotron radiation [25].

If half a wavelength of such a pattern is rotated parallel to an offset axis, permanent magnets with axial fields can be generated (cf. Fig. 2c). However, the end effects will be strong and have to be compensated e.g. by material thickness [26, 27]. Halbach then later generalized this concept to cylindrical [28] and spherical [29] arrangements to create multipolar magnetic fields. Due to his pioneering work in this area, these arrangements are usually all named after him. The pattern shown in Figs. 1 and 2 will typically be referred to as a 1D, linear or planar Halbach array.

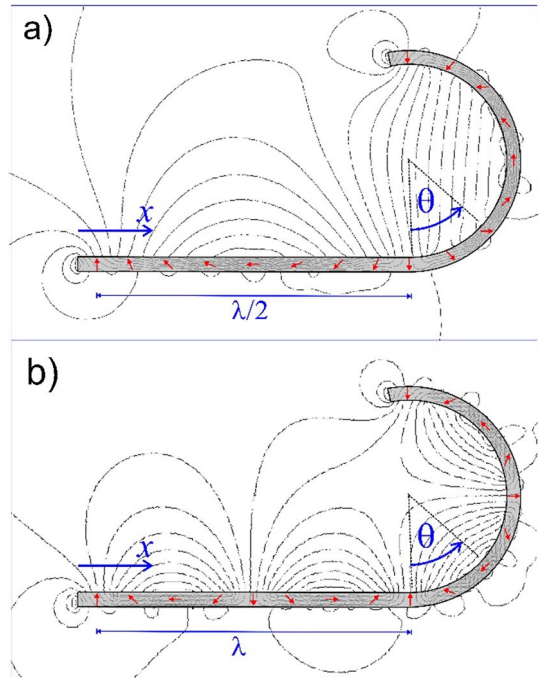

## 3 Cylindrical Halbach Multipoles

If a planar Halbach array of length $\lambda$ is now wrapped around a cylinder like illustrated in Fig. 3a, the variable $x$ in Eq. (1) needs to be replaced by a position angle $\theta = 2\pi\, x/\lambda$ or $x = \theta\, \lambda/(2\pi)$. However, the fact that the magnetization is now lying on a cylinder at an angle $\theta$ must be taken into account by adding this angle to the magnetization position (cf. Fig. 3a). Hence, the magnetization of a cylindrical magnet with one wavelength, i.e. $\boldsymbol{M}(\theta) = M_0\,(\sin 2\theta,\ \cos 2\theta)$, yields a homogeneous (dipolar) field. If two wavelengths of a planar Halbach array are bent to a cylinder (cf. Fig. 3b) the same arguments then result in $\boldsymbol{M}(\theta) = M_0\,(\sin 3\theta,\ \cos 3\theta)$ and so forth.

If this concept is generalized in rolling planar Halbach arrays of length $k\lambda$ up this way, multipolar fields of polarity $p = 2|k|$ are produced (dipole for $|k| = 1$, quadrupole for $|k| = 2$, etc.). The basic design idea is illustrated in Fig. 4a, where the coordinate system has been changed in order to have $\theta = 0$ along the $x$-axis. It is still an idealized magnet with the shape of a hollow cylinder or ring of infinite length and continuously varying magnetization direction. The angle of the magnetization direction, $\varphi$, depends on the position angle, $\theta$, (cf. Fig. 4a) as

$$\varphi = (k+1)\,\theta \quad \text{with} \quad k \in \mathbb{Z}. \tag{3}$$

Figure 4b–g illustrate how the modulus of $k$ determines the polarity, $p = 2|k|$, and its sign the location of the produced magnetic field. For $k > 0$ the field is exclusively

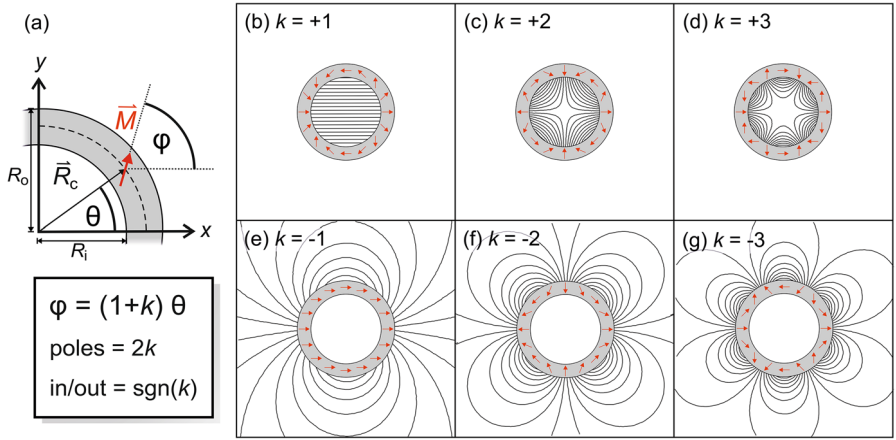

**Fig. 4** Conceptual construction of cylindrical ideal Halbach multipoles: **a** the magnet (gray) consists of a hollow cylinder. Its magnetization $M$ (red arrow) continuously changes with position $R_c$ (defined on a central circle with radius $|R_c| = R_c = (R_o + R_i)/2)$. If the position makes an angle $\theta$ with the x-axis, $M$ is rotated by an angle $\varphi$. This angle $\varphi$ is an integer multiple of $\theta$ depending on the polarity and (inside/outside) location of the magnetic field of the final arrangement (Eq. (3)) as given by the index $k$. The polarity of such rings is then increased from left to right: **b**, **e** display dipolar ($|k| = 1$), **c**, **f** quadrupolar ($|k| = 2$), and **d**, **g** hexapolar fields ($|k| = 3$). The magnetic field is completely inside the cylinder for $k > 0$, i.e. in **b**, **c**, **d** and zero outside. The opposite is the case for $k < 0$. For $k = 0$ the magnetization has radial orientation and no transverse field (only axial, z-direction) field is produced then (color figure online)

inside the ring (as shown in Fig. 3) and for $k < 0$ exclusively on its outside (that would be the case if in Fig. 3 the planar array is bent down and rolled up). The magnetization $M$ [A/m] of such a hollow cylinder changes with respect to the position angle $\theta$ as

$$
\begin{aligned}
M(r) &= |M| \exp{(i\,\varphi)} = |M| \exp{(i[k+1]\theta)} \\
&= \frac{B_R}{\mu_0} \exp{(i[k+1]\theta)} = \frac{B_R}{\mu_0} \begin{bmatrix} \cos{([k+1]\theta)} \\ \sin{([k+1]\theta)} \end{bmatrix},
\end{aligned} \tag{4}
$$

$$
\text{with } r = r \, \exp(i\theta) = r \begin{bmatrix} \cos\theta \\ \sin\theta \end{bmatrix}.
$$

Here, the vectors in the magnet plane are represented in complex and cylinder coordinates, because both are used in the literature. It is also more practical to use the remanent flux density (or remanence) of a permanent magnet instead of its magnetization $M = B_R/\mu_0$ with $\mu_0 \approx 4\pi \cdot 10^{-7}$ kg m/(A s)$^2$, the permeability of vacuum.

For this work only magnetic structures are relevant which encase the magnetic flux density, $B$, they produce. This limits the discussion to $k > 0$ with the following general expression (the asterisk indicates the complex conjugate) (Eq. (21) in [28])

$$
B^*(r) = B_x(r) - iB_y(r) = \begin{bmatrix} B_x(r) \\ -B_y(r) \end{bmatrix} = f(k) \, r^{k-1} \quad \text{with} \quad r = x + iy = \begin{bmatrix} x \\ y \end{bmatrix}. \tag{5}
$$

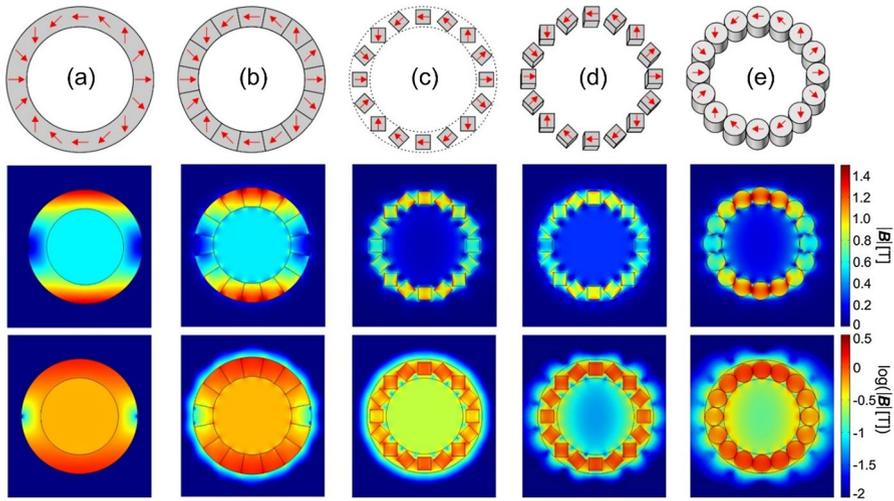

**Fig. 5** Illustration of discretizing the ideal Halbach ring in **a** which has continuously changing magnetization and is infinitely long, into a real system. **b** First the magnet is discretized into $N$ cylinder-segments with a single homogeneous magnetization direction (here $N = 16$). **c** The cylinder segments can be replaced by identical magnets (here squares) which are rotated to the appropriate magnetization direction. **d** Finally, the arrangement is truncated to finite length. **e** Shows another version of **d** using cylindrical magnets. In the top row a sketch of the geometry is displayed, while the central row shows FEM simulations for $B_R = 1.4$ T, the length of the $N$ magnets in **d** and **e** is twice the size. The bottom row is identical to the central one but using a logarithmic color scale to improve the visibility of the inner field distribution (color figure online)

This results in multipolar magnetic fields of amplitude $f(k)$, which is given by

$$f(k) \equiv f^{\text{ideal}}(k) = \begin{cases} B_R \ln \frac{R_o}{R_i} & \text{for } k = 1 \\ B_R \frac{k}{k-1} \left( \frac{1}{R_i^{k-1}} - \frac{1}{R_o^{k-1}} \right) & \text{for } k > 1 \end{cases}, \quad (6)$$

where $R_i$ and $R_o$ are the inner and the outer radius of the hollow cylinder (cf. Fig. 4a), respectively. This amplitude $f(k)$ is the $(k-1)^{\text{th}}$ derivative of the generated flux, which is constant for each polarity. Hence, an ideal dipole ($k = 1$) generates a perfectly homogeneous magnetic field and a quadrupole ($k = 2$) perfectly homogeneous magnetic field gradients, which are orthogonal and have opposite signs.

The fact, that the flux of Halbach rings ($k \geq 1$) is concentrated completely inside, makes them extremely efficient magnets. The efficiency of magnets is defined as the ratio of the energy stored in the accessible region to the maximally storable energy in the magnetic material [30, 31]. Halbach magnets reach a maximum efficiency for dipoles with $R_o = 2.21 R_i$.

These equations are only valid for ideal Halbach rings, which implies that they are continuously magnetized and infinitely long. Therefore, several steps have to be undertaken to convert these ideal Halbach multipoles to more realistic structures (cf. Fig. [5]). This will be included in the Eq. ([5]) of the ideal system by modifying the amplitude of the generated magnetic flux by additional factors, $f$. Equation ([5]) describes the entire spatial flux distribution of the magnetic structure, which will be altered by discretizing and truncating it. Depending on the number of discrete parts, their shapes and distances, this will result in much more complicated field patterns, which cannot be generalized. The following simple equations will only be an estimate of the central field amplitude and say nothing about deviations at larger distances, i.e. about homogeneity. However, they are extremely useful at the planning stage of a magnet design and will give a good estimate about the achievable field/gradient strength.

## 3.1 Permeability

Permanent magnets are usually characterized by their $B(H)$ dependence in the second quadrant, where the most prominent features are their remanent induction $B(0) = B_R$, and the coercitivity field strength $H_c$, for which $B(H_c) = 0$. Typical values in the case of NdFeB permanent magnets are $B_R = 1.3$ T and $H_c = 1$ MA/m. The relative permeability of the magnetic material $\mu_r$ is described as the slope of this line, i.e. $\mu_r = \partial B(H)/\partial H$.

Usually, the structural materials in the construction are diamagnetic or slightly paramagnetic (aluminum, copper) and can be described by the linear relationship $B = \mu_0 \mu_r H$, with values of $\mu_r$ differing from unity by only about $10^{-5}$ (which can be also expressed as, $\mu_r = 1 + \chi$, with $\chi$ the magnetic susceptibility). For this reason, these materials can be neglected compared to permanent magnets ("neodymium magnets" or $Nd_2Fe_{14}B$ typically have $\mu_r = 1.05$, and "samarium-cobalt magnets" $SmCo_5$ or $Sm_2Co_{17}$ have $\mu_r = 1.03$–$1.11$, cf. Appendix A). In order to estimate the effect of these not quite negligible permeability values we treat a Halbach ring like a continuous shell which shields the field of a larger ring surrounding it. Thus, an ideal Halbach ring can be treated like a cylindrical shield with an inner field reduced by a factor (Eq. (72) in [32])

$$f^\mu = \frac{4\mu_r}{\left(\mu_r + 1\right)^2 - \left(\mu_r - 1\right)^2 \left(\frac{R_i}{R_o}\right)^2}. \tag{7}$$

If we imagine the magnetized hollow cylinder as a superposition of very thin cylindrical shells, the field generated by the outermost shell is shielded by all the inner shells, while the innermost is only shielded by the sample volume. The solution for this scenario is given by (Eq. (31) in [33])

$$f^\mu = \frac{(1 - \Lambda)\Lambda R_o^2}{R_i^2 - \Lambda^2 R_o^2} \quad \text{with} \quad \Lambda \equiv \frac{\mu_r + 1}{\mu_r - 1} \quad \text{and} \quad \mu_r \neq 1. \tag{8}$$

Simulations for the relevant range of the permeability show that Eq. ([8]) can be approximated by (with an error smaller than 1% for < 1.35)

$$f^\mu = \frac{1}{\sqrt{\mu_r}}. \tag{9}$$

If more than one Halbach ring is used to construct a (nested) system, the shielding of the field originating from the outer rings by the permeability of the inner rings must also be taken into account (e.g. by using Eq. (7)).

## 3.2 Segmented Rings

Unlike for planar Halbach arrays there is no accurate technique available yet to magnetize a hollow cylinder of permanent magnet material with the desired continuous magnetization patterns as demanded by Eq. (4). A real Halbach ring is therefore always an approximation of the ideal Halbach multipole by constructing it from $N$ segments each with a single magnetization direction (cf. Fig. 5b). This type of segmentation reduces the inner field by (Eq. (24b) in [28])

$$f^{seg}(k) = \frac{\sin((k+1)\,\pi/N)}{(k+1)\,\pi/N}. \tag{10}$$

This correction becomes mainly important for small $N$ and high $k$ (e.g. for $k=1$ and $N=8$, $f^{seg}=0.6366$, but for $N=16$, $f^{seg}=0.9745$). It is trivial that the level of discretization must be at least the number of poles ($N \geq 2\,k$), an equivalent of Nyquist's theorem of "sampling" the theoretical magnetization wave that defines the multipole (cf. Figs. 2 and 3). At this limit the magnetization vectors of neighboring segments point in opposite radial directions.

## 3.3 Non-cylindrical Segments

Replacing the cylindrical segments of Fig. 5b by $N$ identical magnets with a polygonal, round or trapezoidal footprint (as shown in Fig. 5c exemplarily by squares) can be advantageous in terms of costs but also for strategies to optimize homogeneity (see Sect. 4.1). This concept is also named Mandhalas [34] and the pros and cons of differently shaped sub-pieces are discussed in [35]. Obviously, the reduction of magnetic volume by this approach reduces the magnetic flux density additionally. If the footprint of the sub-pieces covers an area $A_M$, the strength of the resulting magnet is reduced by the ratio to the area of the ideal Halbach (i.e. $\pi\left(R_o^2 - R_i^2\right)$).

$$f^M = \frac{N\,A_M}{\pi\left(R_o^2 - R_i^2\right)}. \tag{11}$$

The geometry (calculation of vertices and distances) of dense arrangements of polygonal pieces is described in Appendix B.

### 3.4 Finite Length

So far, the magnets were described only in two dimensions, corresponding to an infinite length in the third or $z$-dimension. Obviously, a real magnet must be truncated to a certain length, $L$. Zijlstra gives an analytical expression (Eq. (49) in [36]) for the reduction factor, $f^L(k)$, of the field in the center of such a truncated Halbach dipole

$$f^L(1) = 1 \; + \; \frac{1}{\ln\left(R_o/R_i\right)}\left(\frac{L}{2I} \; - \; \frac{L}{2O} \; - \; \ln\frac{L+O}{L+I}\right)$$
$$\text{with} \;\; I \; = \; \sqrt{L^2 + 4R_i^2} \;\; \text{and} \;\; O \; = \; \sqrt{L^2 + 4R_o^2}. \tag{12}$$

In order to generalize this for multipolar Halbach rings, they are constructed by a continuous distribution of dipoles on the circumference of a circle with radius $R_c$. Then the decay along the third dimension (for $k = 1$ see Eq. (8) in [37]) is given by

$$B_{xy}(z) \; = \; \frac{R_c^{2k+3}}{\left(R_c^2 + \left(z - z_0\right)^2\right)^{k+3/2}} \; B_{xy}(0) \quad \text{for} \;\; k \; \geq \; 1$$
$$\text{and for} \;\; x, y \; = \; 0 \quad \text{with} \quad R_c \; = \; \frac{R_i + R_o}{2}. \tag{13}$$

In order to account for the length, $L$, of the magnet, Eq. (13) is integrated over $z_0 = \pm L/2$ and set in relation to the integral for the infinitely long ideal cylinder ($z_0 = \pm \infty$). The reduction factor, $f^L$, is then found for $z = 0$. For the two relevant cases (i.e., $k = 1$ and $k = 2$) the following simple expressions can be found (for $k = 1$ see Eq. (A.5) in [23]) at $z = 0$.

$$f^L(k) = \frac{\displaystyle\int\limits_{-L/2}^{L/2} B_{xy}(k, z = 0, z_0)\, dz_0}{\displaystyle\int\limits_{-\infty}^{\infty} B_{xy}(k, z = 0, z_0)\, dz_0} \; = \; \begin{cases} \frac{L(L^2 + 6R_c^2)}{\left(L^2 + 4R_c^2\right)^{3/2}} & \text{for} \;\; k \; = \; 1 \\[2mm] \frac{L(L^4 + 10L^2R_c^2 + 30R_c^4)}{\left(L^2 + 4R_c^2\right)^{5/2}} & \text{for} \;\; k \; = \; 2 \end{cases}. \tag{14}$$

The solutions for higher $k$ values and a general equation are listed in Appendix C. Note that it is still the field components in the $xy$-plane whose decay in $z$-direction is described here. The expression in Eq. (14) for $k = 1$ can be found from Eq. (12) for a very thin ring by expressing $R_o = R_c + \delta r$ and $R_i = R_c - \delta r$ as the zeroth order expansion in $\delta r$.

In Sect. 4.3 it is shown how the decay $\sim z^{-5}$ for $k = 1$ along the central axis can be avoided by using multiple rings with gaps in between them [37].

As already mentioned in the introduction, for most magnetic resonance experiments dipolar ($k = +1$) and quadrupolar ($k = +2$) Halbach magnets are of interest only. Therefore, the effects discussed in the previous sections can be summarized as

Dipole: ($k = 1$)

$$\boldsymbol{B}_{\mathrm{D}}(x, y, 0) = \begin{pmatrix} B_x \\ 0 \\ 0 \end{pmatrix} \quad \text{with} \quad B_x = B_{\mathrm{R}} \ln \frac{R_0}{R_{\mathrm{i}}} f^{\mu} f^{\mathrm{seg}}(1) f^{\mathrm{M}} f^{\mathrm{L}}(1)$$

$$B_x(x, y, 0) = \frac{B_{\mathrm{R}}}{\sqrt{\mu_{\mathrm{r}}}} \ln \frac{R_{\mathrm{o}}}{R_{\mathrm{i}}} \frac{\sin 2\pi/N}{2\pi/N} \frac{N\, A_{\mathrm{M}}}{\pi \left(R_{\mathrm{o}}^2 - R_{\mathrm{i}}^2\right)} \frac{L(6R_{\mathrm{c}}^2 + L^2)}{\left(4R_{\mathrm{c}}^2 + L^2\right)^{3/2}}.$$

(15)

Quadrupole: ($k = 2$)

$$\boldsymbol{B}_{\mathrm{Q}}(x, y, 0) = \begin{pmatrix} B_x \\ -B_y \\ 0 \end{pmatrix} \quad \text{with} \quad B_x = G_{\mathrm{Q}} x \quad \text{and} \quad B_y = G_{\mathrm{Q}} y$$

$$G_{\mathrm{Q}} = \frac{2B_{\mathrm{R}}}{\sqrt{\mu_{\mathrm{r}}}} \left( \frac{1}{R_{\mathrm{i}}} - \frac{1}{R_{\mathrm{o}}} \right) \frac{\sin(3\pi/N)}{3\pi/N} \frac{N\, A_{\mathrm{M}}}{\pi \left(R_{\mathrm{o}}^2 - R_{\mathrm{i}}^2\right)} \frac{L(L^4 + 10L^2 R_{\mathrm{c}}^2 + 30R_{\mathrm{c}}^4)}{\left(4R_{\mathrm{c}}^2 + L^2\right)^{5/2}}.$$

(16)

Equations (15) and (16) are very helpful for first estimations of field strengths of magnet geometries. A table in Appendix D compares the quality of the terms with the more accurate results from FEM simulations. The agreement is very good (discrepancies are less than a percent, except for the last two factors).

### 3.5 Demagnetizing Field

If a certain geometry is found that produces the desired magnetic flux density, $\boldsymbol{B}$, inside the magnet system one has to check if the chosen material is magnetically hard enough. Inside the magnetic material the magnetic field, $\boldsymbol{H}$, is not in the same direction as the magnetization, $\boldsymbol{M}$. If in certain regions the component of $\boldsymbol{H}$ in the direction opposite to $\boldsymbol{M}$ exceeds the coercivity, $H_{\mathrm{c}}$, the local magnetization will be quenched or demagnetized (for Halbach systems [31, 38]). Otherwise, permanent magnets could be built that generate virtually any field strength by adding more magnetic material. Hence, simulations that account for the full $BH$- or $MH$-curve are needed. Alternatively, calculations of the magnetic field inside the magnetic material can be inspected if the coercivity is exceeded (typical values and their dependence on temperature are listed in Appendix A). The intrinsic coercivity, $H_{\mathrm{c}}$, is the magnetic field, which is needed to reduce the magnetization to zero, but the magnetization starts to deviate from its initial slope even earlier. However, in rare-earth magnets this transition is very steep. The magnetic field, $\boldsymbol{H}$ (see Fig. 6c), can be calculated from the flux density, $\boldsymbol{B}$ (see Fig. 6b), and the magnetization of the material, $\boldsymbol{M}$ (cf. Eq. (4) and Fig. 6a)

$$\boldsymbol{H} = \frac{\boldsymbol{B}}{\mu_0} - \boldsymbol{M} \quad \text{and} \quad |\boldsymbol{M}| = \frac{B_{\mathrm{R}}}{\mu_0}.$$

(17)

Figure 6 illustrates this effect on a thick-walled Halbach dipole and quadrupole. The internal magnetic field exceeds the coercivity of the chosen material in several

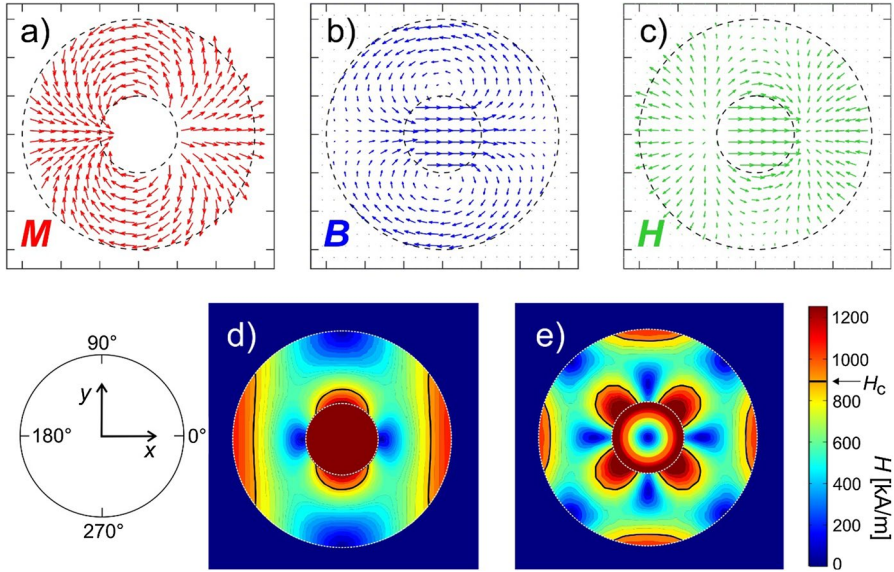

**Fig. 6** Illustration of the effect of demagnetization on a thick-walled Halbach dipole (**a**–**d**) and a quadrupole (**e**) with identical dimensions (dashed white lines). **a** The magnetization as requested by Eq. (4), **b** shows the magnetic flux density produced by **a**. **c** Combining $M$ and $B$ according to Eq. (17) gives the magnetic field $H$. **d** Images of the magnitude $H$ for the same dipole and **e** for a quadrupole. The material properties were chosen from the FeNdB-material N54 with $B_R = 1.47$ T, $H_c = 880$ kA/m (black lines in **d/e**) in Appendix A, $R_o/R_i = 3$. A coordinate system is shown on the lower left (color figure online)

regions (marked by black lines in Fig. 6d, e) at the outer and inner rim, which would be demagnetized and hence no longer contribute to the planned magnetization pattern, causing a lower field and homogeneity than predicted.

In [31] Bjørk et al. calculated conditions for these inner and outer regions with highest magnetic field. Demagnetization is avoided if

$$\text{inner rim:} \quad H_c \; > \; \frac{B_R}{\mu_0} \ln \frac{R_o}{R_i} \quad \text{for } k = 1 \quad \text{and} \quad H_c \; > \; \frac{k}{k-1} \frac{B_R}{\mu_0} \quad \text{for } k > 1$$

$$\text{outer rim:} \quad H_c \; > \; \frac{B_R}{\mu_0}.$$

(18)

At the outer rim these regions are located at the poles (0° and 180° for a dipole) and at the inner rim between them (90° and 270° for a dipole).

However, it may be worth noting that for accelerators Halbach dipoles ($R_i = 3$ mm, $R_o = 100$ mm) have been constructed which reach almost 4 T (homogeneity 0.4%) [17] and recently even 5.16 T in a 2 mm gap [39]. The same group also developed extremely strong multipoles [40], e.g. quadrupoles with gradients of ca. 300 T/m [39]. This was achieved by using various materials with different $H_c$ to avoid demagnetization.

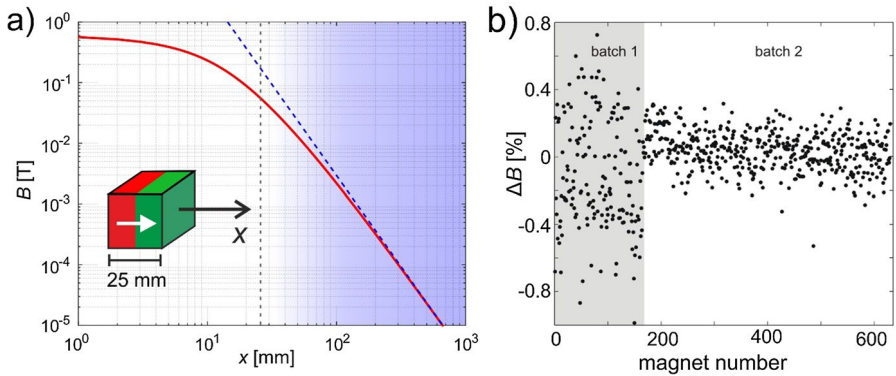

**Fig. 7** **a** dipole approximation for characterizing magnets by their far field: double log-plot of the magnetic field, $B$, (red line) produced outside a cube-shaped magnet in magnetization direction, $x$. The blue dashed line shows the expected slope for a dipole ($x^{-3}$). The gray dashed line indicates the size of the magnet. The purple shaded region suggests a region for measurements of "dipole-like" behavior. The $B$-field of the cubic magnet was simulated for $B_R = 1.4$ T, hence corresponds to a magnetic moment $m = B_R V/\mu_0 = 17.4$ Am$^2$. Note that an ideal spherical magnet of the same volume would generate a dipolar field everywhere beyond its surface. **b** Result of such measurements: The individual deviation, $\Delta B$, of 620 FeNdB-magnets from the mean. The numbering was provided by the producer. The shaded regions identify two different batches by significantly different variances of unknown origin (color figure online)

Finally, one should be aware that $H_c$ is also strongly temperature dependent (cf. Table 1).

# 4 Homogenizing

## 4.1 Homogenizing in the Magnet Plane

The magnetic fields of ideal Halbach multipoles are flawless (see Figs. 5a, 6d, e). The inhomogeneities introduced by discretization increase with each step in Fig. 5. At first glance, it is unclear how the concept of Mandhalas [34] (constructing the magnet from the same magnetic parts) might help here. To understand this strategy the main sources which cause field inhomogeneities have to be discussed. Typically, the production and magnetization process of the permanent magnets is outsourced to companies that have specialized in this field. Usually, the quality of these building blocks of the Halbach system is most crucial for the final result, and unfortunately the underlying steps are out of direct control. Figure 7b shows the variation of the field produced at a fixed distance from the surface of some 600 permanent magnets. The variation can easily reach percent ranges, but clearly the numbering provided by the producer reveals two batches in the production by the change in variance. Such variations can have their origin in the production of the magnets and/or the magnetization process.

Another problem arises from the accuracy of the magnetization direction. The cylindrical segments in Fig. 5b for instance need correct individual magnetization angles. It is easy to imagine that this will add another error in the percent range. An easy strategy to circumvent these problems is to order magnets of identical shape and measure their far field. Figure 7a shows how the magnetic field drops off the surface of a cube-shaped magnet in this example. At a distance roughly equal the size of the magnet, a dipolar regime is entered (dashed blue line), which means that shape and local magnetization variations inside the magnet average out and the magnet can be regarded as a spherical magnet of the same volume, or equivalently a point source described as a magnetic dipole. Note that this distance should be not too far away to ensure sufficient accuracy in measuring its stray field. In Appendix E a simple method is described that uses this dipole approach to minimize the error in the center of the arrangement [37]. This strategy typically improves the homogeneity by an order of magnitude. However, this strongly depends on the number of magnets to select from, their deviation and the chosen geometry. More evolved methods optimize larger field maps for different arrangements [6, 41, 42].

Using magnets with a polygonal footprint further reduces the error in misalignments of the magnetization direction because it coincides with a certain edge. However, edges in the magnet shape introduce higher polarities in the resulting field (cf. Fig. 5c, d), which can be reduced by using round magnets (spheres, bullet shaped, cylinders, cf. Fig. 5e). If one uses such round magnets, their correct orientation can be achieved by using a magnetic alignment procedure [35] giving excellent homogeneities. All the described concepts assume that one really uses only the innermost section of the magnet (somewhat like the far field). If the sample space is close to the size of the opening ($2R_i$) of the ring, the spatial inhomogeneity of the individual magnets can no longer be neglected (cf. Fig. 10a).

For dipoles Tewari et al. showed that the rotation angle $\varphi_i = (k+1)\,\theta_i$ (see Fig. 4a) of each magnet is not necessarily the best choice for optimal homogeneity [43]. From the same group there is also the suggestion to deviate from a circular arrangement and that arranging magnets on ellipses instead also improves overall homogeneity [44].

## 4.2 Axial Homogenization

The truncation of an ideal Halbach multipole to a certain length or height introduces an additional inhomogeneity of $B_{xy}$ along the $z$-axis. This decay can be approximated by Eq. (14), and as already stated in Sect. 3.4 it is advantageous to separate the Halbach cylinders into differently spaced rings. This approach is very much like in Helmholtz coils or solenoids with denser pitch towards their ends to improve the homogeneity along their axis [37]. To homogenize the central field, $B_{xy}^{\Sigma}$, produced by two dipole rings placed at distances $\pm s_1$, the second derivative has to become zero at the center $z = 0$.

$$B_{xy}^{\Sigma}(0,0,z) = R_c^5 B_{xy}(0,0,0) \left[ \frac{1}{\left(R_c^2 + (z+s_1)^2\right)^{5/2}} + \frac{1}{\left(R_c^2 + (z-s_1)^2\right)^{5/2}} \right],$$

(19)

with the second derivative

$$\frac{\partial^2 B_{xy}^{\Sigma}(z)}{\partial z^2} = 5R_c^5 B_{xy}(0,0,0) \left[ \frac{6(z+s_1)^2 - R_c^2}{\left(R_c^2 + (z+s_1)^2\right)^{9/2}} + \frac{6(z-s_1)^2 - R_c^2}{\left(R_c^2 + (z-s_1)^2\right)^{9/2}} \right].$$

(20)

In the center ($z=0$) this has to become zero for a flat maximum (cf. Fig. 8a)

$$\frac{\partial^2 B_{xy}^{\Sigma}(0)}{\partial z^2} = 0 \quad \Rightarrow \quad s_1 = \pm\frac{R_c}{\sqrt{6}} \qquad \text{and}$$

$$B_{xy}^{\Sigma}(0,0,0) = \frac{72}{49}\sqrt{\frac{6}{7}}\, B_{xy}(0,0,0).$$

(21)

Generally, this is fulfilled at $s_1 = \pm R_c/\sqrt{2k+4}$ for $k \geq 1$. Appendix F lists the optimal distances for up to 12 rings (cf. Fig. 8c) for $k=1$—6.

Figure 8b shows a differential arrangement, which generates a constant gradient $\partial B_{xy}^{\Delta}/\partial z$. For this the optimal distance is $s_1 = \pm R_c/\sqrt{2}$ with a gradient of $\frac{\partial B_{xy}^{\Delta}}{\partial z} = \frac{80}{81}\sqrt{3}\, \frac{B_{xy}(0,0,0)}{R_c}$ in the center.

Again, these distances are first starting points, which need refinement via simulations. The very homogeneous regions in Fig. 8a, c are, of course, limited to the axis. Farther away from the axis the gaps between the magnets will cause inhomogeneities increasing with radial distance (cf. Fig. 10b). Strategies to remove them will be discussed in the next section. An alternative to stacks of rings of the same length is to change the thickness of the individual rings [45] and remove/reduce the gaps in between them as shown in Fig. 8d. Nevertheless, the use of only one magnet size has to be sacrificed for this approach and the presented solutions in Appendix E are no longer valid because they were obtained for infinitesimal thin rings. However, the density distribution presented by them can be used as a starting point to optimize the ring thicknesses.

Another possibility to build Halbach arrays without end effects by truncation would be to use Halbach spheres [18]. This is more of a theoretical concept because the authors are not aware that anyone ever built such a demanding structure for magnetic resonance (in [46] a sphere has been roughly approximated). Figure 9a shows the principle layout of such a Halbach sphere generated by rotating the cross section of an ideal Halbach dipole around the axis of its poles [29, 47]. A Halbach sphere has 4/3 more internal flux than an ideal ring with the same shell thickness and additionally a strong stray field. At first glance it looks a bit foolish to wall an apparatus inside a hollow sphere, but there are certain angles where such spheres can be opened without force [1] to provide access to the inner volume (see also Sect. 5.1). This inner space can be extended by using Halbach hemispheres as endcaps of a Halbach ring (cf. Fig. 9b) [48]. The field strength of such a hemisphere is then half of that of the complete sphere, and the Halbach cylinder outer radius has to be adopted to match this by being $R_o^{2/3}R_i^{1/3}$ of the hemispheres. Again, no one, to our knowledge, has ever built such a device.

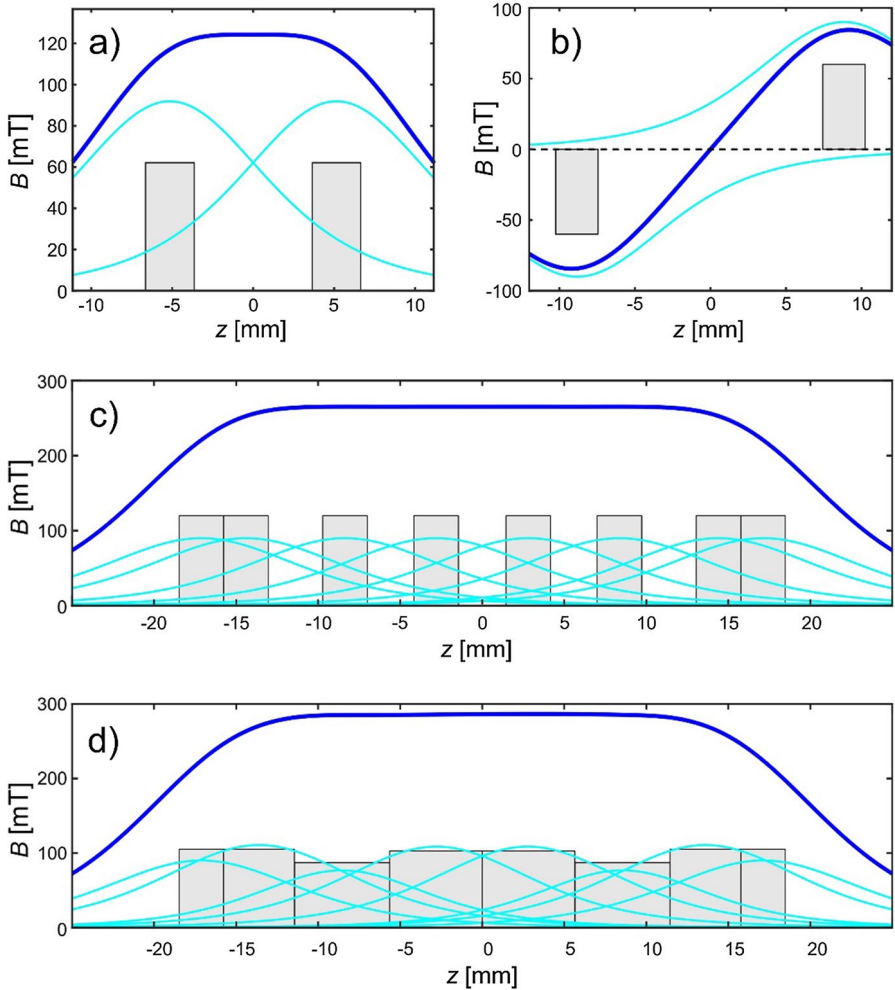

**Fig. 8** Illustration of stacking Halbach multipoles with optimized gaps along their axis. Dark blue: resulting combined magnetic field, $B_{xy}^{\Sigma}$, of the arrangement. Cyan: magnetic field of each individual Halbach cylinder (gray rectangle). **a** Two dipole cylinders at the distance calculated in Eq. (21), **b** two antiparallel dipoles generating a constant gradient of $B_{xy}$ along $z$. **c** 8 stacked rings (see Appendix E). For all: $k = 1$, $R_c = 12.5$ mm, $L = 3$ mm, $B_R = 1.4$ T. **d** Instead of identical rings stacked at different distances one can also use rings of various $L$ and different thickness (here $R_i$ was kept constant and $R_o$ adapted to homogenize the field) (color figure online)

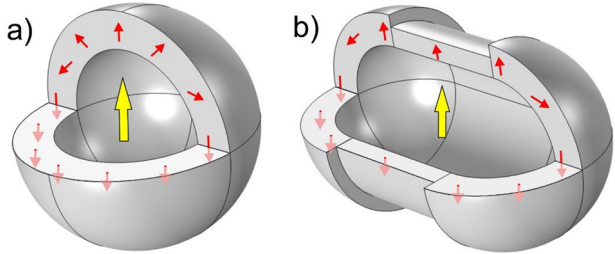

**Fig. 9** **a** Halbach dipole sphere, **b** using Halbach hemispheres as endcaps of a Halbach cylinder. In both drawings a quarter is cut away. Red arrows sketch the magnetization direction and the big yellow arrow the resulting flux (color figure online)

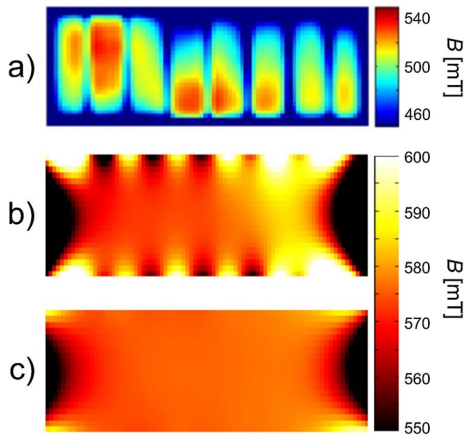

**Fig. 10** **a** Normal field component at a distance of ca. 1 mm over the pole surface of 8 FeNdB-magnets stacked together using aluminum spacers between them. They are all from the same batch of magnets and should be identical. **b** Magnetic flux measured over a similar arrangement but now being part of a Halbach magnet [4] and measured at a different distance, which explains the different field strengths. **c** Same as **b** but all magnet poles covered with 1 mm thick iron sheets. Please note the different color scales in **a** and **b/c** (color figure online)

## 4.3 Passive Shimming by Iron Pole Plates

Figure 10a shows the magnetic flux density measured close to the pole surface of eight magnets. Quite strikingly each of them shows a pronounced flux distribution over its surface. Although the methods described in Sect. 4.1 can compensate their overall strength at a distance, there are situations where one has to use as much of the inner volume of a magnet as possible. In [4] it was found that flat sheets of iron almost completely cancel this inhomogeneity out (cf. Fig. 10b, c). The iron plates (or other soft magnetic material with high permeability) act like a pole piece, and by bringing the various spatial magnetic field contributions to a similar magnetic potential smooths out the biggest inhomogeneities. Below about 1.5 T saturation effects are not expected in iron or low-carbon steels, but as it clearly can be seen in Fig. 10c, they also reduce the magnetic flux.

Recently, it has been reported that magnetic filaments for 3D-printers can be used to easily create arbitrarily shaped pole pieces or be already embedded in the support for magnets [49], however their low permeability ($\mu_r < 10$) may become a problem.

## 4.4 Mechanical Shimming

Due to the variation in the magnetic materials, production and positioning tolerances it is hard to imagine building Halbach magnets with homogeneities much less than 100 ppm deviation of the magnetic flux over a larger part of the inner accessible volume, which is a comparable value for superconducting magnets before cryoshimming. Typically, the last mechanical correction steps have to take spatial field information into account to find the optimal configuration for an individual magnet (a step which may already require the incorporation of ambient field distortions).

For permanent magnets this correction can be done by individually position either additional magnets or materials of high permeability [50]. Such a semi-permanent correction of course is best planned from detailed spatially resolved field maps. The corrective material can be inserted at suitable positions and/or mechanically moved (e.g. by set screws) for fine adjustments [51–53] or being rotated [54–56]. One major aspect is that not too much of the available sample space should be sacrificed by this additional equipment. A very elegant and space preserving concept is shown in Fig. 11 [57]. The very small magnet produces a field of 0.7 T and allows simple spectroscopy shown on the [1]H-spectrum of toluene acquired in a 5 mm standard NMR tube (inset Fig. 11a). The resolution measured for a water sample was 4.5 Hz or 0.15 ppm at half height. This homogeneity was achieved first by shifting the rectangular sections to suitable positions and finally using additional coils for finer field adjustment of the first-order terms only. In the meantime this idea has been further developed and successfully commercialized [58].

Another idea for simple, adjustable and stationary shims is presented in Fig. 12. It makes use of a pair of permanent magnets both with the shape of a hollow cylinder and transverse magnetization, which are aligned coaxially (cf. Fig. 12 top row). They will have the same stray field if both have the same height and their geometry and remanences are chosen such that they both have the same product of remanence times their area, i.e.

$$B_R^A \, \pi \left( R_{Ao}^2 - R_{Ai}^2 \right) = B_R^B \, \pi \left( R_{Bo}^2 - R_{Bi}^2 \right), \tag{22}$$

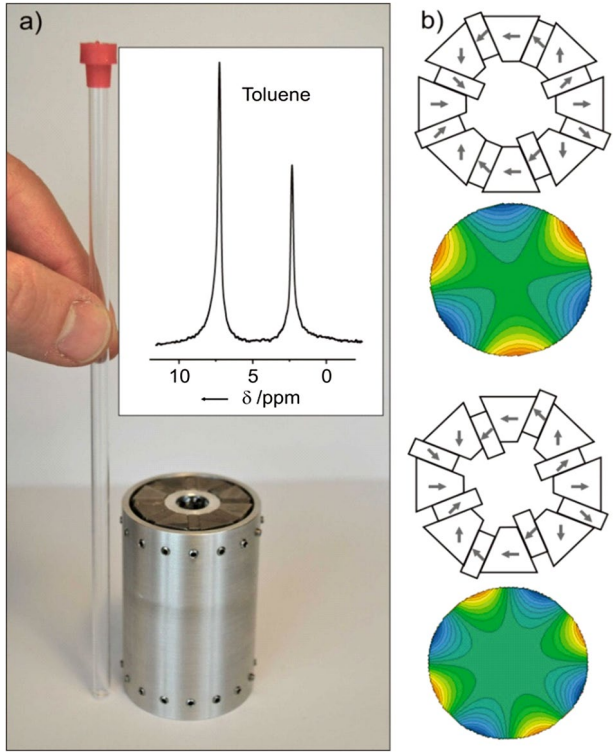

**Fig. 11** Halbach dipole built from fixed trapezoidal and movable rectangular units: **a** Picture of the 0.7 T magnet next to a standard 5 mm tube for the liquid sample. Inserted is the $^1$H-NMR spectrum of toluene of the homogenized system. **b** Demonstration how the movement of the rectangular magnets can generate magnetic fields of third order (top) and fourth order (bottom). Below each magnet arrangement is the resulting field map to illustrate the effect. Reproduced from [57] with permission from the Royal Society of Chemistry (color figure online)

where A/B indicates the inner or outer cylinder ($R_{Ao/i}$ = outer/inner radius of cylinder A). Their fields add up when both point in the same direction. Their vector sum is minimal (vanishes if Eq. (22) is fulfilled) for antiparallel orientation as illustrated in Fig. 12. If a set of $K$ such units is mounted on a circle with radius $R_s$ and their geometry is chosen small compared to the rest of the magnet system, their influence in the center is approximately that of a dipole and the additional flux at a point $\boldsymbol{p} = [x, y]$ close to the center is given by

$$\boldsymbol{B}(\boldsymbol{r}) = \sum_{j=1}^{K} \frac{\mu_0}{4\pi} \frac{3\boldsymbol{r}_j\,(\boldsymbol{m}_j\,\cdot\,\boldsymbol{r}_j)\,-\,\boldsymbol{m}_j r_j^2}{r_j^5} \quad \text{with} \quad \boldsymbol{r}_j = R_s \begin{pmatrix} \cos\theta_j \\ \sin\theta_j \end{pmatrix} \,-\, \boldsymbol{p}$$

$$\text{and} \quad \boldsymbol{m}_j = m_j^A \begin{pmatrix} \cos\varphi_j^A \\ \sin\varphi_j^A \end{pmatrix} + m_j^B \begin{pmatrix} \cos\varphi_j^B \\ \sin\varphi_j^B \end{pmatrix}$$

$$\text{or} \quad \boldsymbol{m}_j = \frac{B_R\,V}{\mu_0} \begin{pmatrix} \cos\varphi_j^A + \cos\varphi_j^B \\ \sin\varphi_j^A + \sin\varphi_j^B \end{pmatrix} \quad \text{for} \quad m_j^A = m_j^B,$$

$$(23)$$

where $\theta_j$ is the angle at which the $j^{\text{th}}$ shim cylinder-pair is positioned, whose outer and inner cylinders are rotated by angles $\varphi_j^A$ and $\varphi_j^B$. The last equation is given if both cylinders with total volume $V$ are made from the same material with remanence $B_R$.

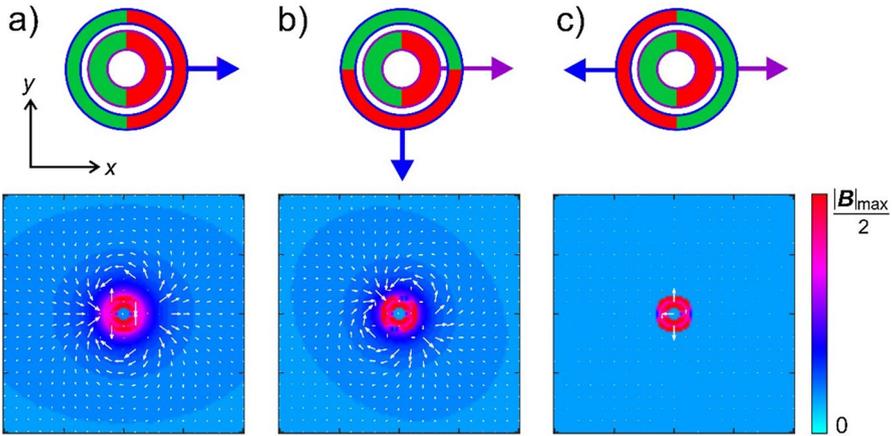

**Fig. 12** Shimming using a pair of hollow cylinders with transverse magnetization (blue and purple arrows in top row). If their size and strength are matched according to Eq. (22) and if both can be rotated around a common concentric axis they produce the flux depicted in the lower row. To illustrate this principle, the inner cylinder stays at the same position and the outer one is rotated from **a** parallel, **b** perpendicular to **c** antiparallel, where the combined field of both vanishes. The field maps in the lower row show the magnitude of the field (colors limited to the lower half of the full range) with an overlaid vector plot, nicely showing the dipole characteristics of the magnetic field (color figure online)

The correction field can then rather simply be calculated by the superposition of $K$ individual dipoles of individual strength and angle to homogenize the original magnetic field. From that the angles of each cylinder pair are determined using Eq. (23). This approach has the advantage that it only uses components on a fixed axis. The method was successfully tested but needs further refinement [59].

If the instrument is going to be used for NMR spectroscopy the final fine adjustment ("active shimming") of homogeneity is typically and reasonably done by a set of coils charged by finely regulated and stabilized currents. However, these coils deviate from the standard shim-coil designs used in superconducting magnets with axial fields. Due

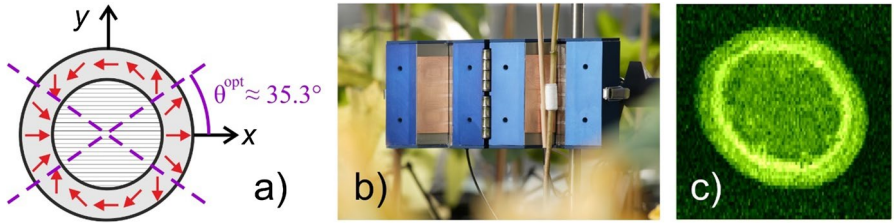

**Fig. 13** **a** Geometry of a Halbach dipole that opens without force at an angle $\theta_0^{\text{opt}} \approx \pm 35.3°$ relative to the poles. **b** Photograph of a prototype (NMR-CUFF) comprising four magnet stacks (of 8 magnets each, cf. Fig. 10, generating a magnetic field of 0.57 T), which opens at $\alpha = 45°$. Furthermore, it is equipped with gradient coils to image a stem of a plant in vivo. The resulting MRI is shown in **c**. For details see [4] (color figure online)

to the transverse field in Halbach magnets, the coil geometry has to be adapted (e.g. using cosine-coils [60]).

## 5 Additional Features

### 5.1 Force-Free Openable Magnets

For some applications it can be useful to open Halbach rings and particularly spheres, e.g. to clamp them around a tube or plant or to insert samples (cf. Fig. 13). Following the discussion in [4] Halbach multipoles can be opened without forces at angles, $\theta_j^{\text{opt}}$, which are solutions to (see Appendix G, Eq. (42))

$$3 \cos(2k\theta) = 1, \tag{24}$$

resulting in

$$\theta_j^{\text{opt}} = \frac{\pm \cos^{-1}(1/3) + 2\pi j}{2k} \quad \text{with } j \in \mathbb{Z}. \tag{25}$$

Again, these are theoretically derived values for a dipole model. However, there are always clear minima in Halbach arrangements, but the optimal position to open such a magnet system depends also on the degree of discretization (see Appendix G). The optimal angle for $k = 1$ is $\theta_0^{\text{opt}} \approx \pm 35.26°$ and for $k = 2$, $\theta_0^{\text{opt}} \approx \pm 17.63°$. This variation of mutual forces can actually be experienced when building a Halbach dipole. Whereas the magnets close to $\theta = 0$ repel each other, they attract each other around $\theta = 90°$.

### 5.2 Nested Magnets with Adjustable Fields via Rotation

One of the most interesting ideas of using Halbach rings is to align them coaxially and rotate them to change the amplitude of the generated magnetic flux density in the center [61, 62]. Due to the fact that Halbach inner multipoles ($k \geq 1$) have no stray fields they are not a magnet from the outside. Consequently, there is only negligible torque if they are rotated in the homogeneous field of another outer dipole [63]. The situation changes for multipoles which produce spatial field changes, but still the torque is relatively small, yet with some unavoidable cogging [33].

If $\boldsymbol{B}$ is a constant vector field over the entire sample region (i.e. a dipole, $k=1$), its rotation by an angle $\alpha$ is given by

$$\boldsymbol{B}' = \mathbf{R}(\alpha)\,\boldsymbol{B} = \begin{pmatrix} \cos\alpha & -\sin\alpha \\ \sin\alpha & \cos\alpha \end{pmatrix} \begin{pmatrix} B_0 \\ 0 \end{pmatrix} = \begin{pmatrix} B_0\cos\alpha \\ B_0\sin\alpha \end{pmatrix}, \qquad (26)$$

where $\boldsymbol{B}'$ is the rotated field by applying the rotation matrix $\mathbf{R}(\alpha)$.

For $k > 1$, however, the magnetic field depends on $\boldsymbol{r}$ (cf. Eq. (5)), and a real vector field rotation must be applied, e.g. for the gradients of a quadrupole ($k=2$, see Eq. (16) and [2, 64])

$$\begin{aligned}
\boldsymbol{B}'_{Q}(\boldsymbol{r}) &= \mathbf{R}(\alpha)\,\boldsymbol{B}_{Q}(\boldsymbol{r}) \cdot \left( \mathbf{R}^{-1}(\alpha)\,\boldsymbol{r} \right) \\
&= \begin{pmatrix} \cos\alpha & -\sin\alpha \\ \sin\alpha & \cos\alpha \end{pmatrix} G_{Q} \begin{pmatrix} 1 & 0 \\ 0 & -1 \end{pmatrix} \cdot \left[ \begin{pmatrix} \cos\alpha & \sin\alpha \\ -\sin\alpha & \cos\alpha \end{pmatrix} \begin{pmatrix} x \\ y \end{pmatrix} \right] \\
&= G_{Q} \begin{pmatrix} x\cos 2\alpha + y\sin 2\alpha \\ x\sin 2\alpha - y\cos 2\alpha \end{pmatrix}.
\end{aligned} \qquad (27)$$

Hence, the gradient rotates by $2\alpha$ when the quadrupole is rotated by an angle $\alpha$ [64] (or generally the $k^{\text{th}}$ derivative of the magnetic field of a Halbach multipole with $k \geq 1$, rotates with $k\alpha$ when the multipole is rotated by $\alpha$).

The great advantage here is the possibility to change field and gradient strength when using nested magnets with the same polarity $k$. If, for instance, two dipoles are combined coaxially with radii chosen such that they produce the same flux, their combined field can be changed from twice that of a single one (parallel dipoles) to zero (antiparallel dipoles) as shown for dipoles and quadrupoles in Fig. 14.

This not only allows to adjust homogeneous fields for DNP [11, 12] or do field sweeps for EPR [9], but can also be used for imaging with gradients adjustable in strength and direction [64] (this will need a dipole and two rotating quadrupoles).

In this way very strong fields and gradients can be changed in less than a second, which may even be an alternative to electromagnets (considering their large inductance). In [65] a counter-rotating pair of dipoles generated an AC field of 150 mT with frequencies of about one Hz, a concept that may be also interesting for diffusion NMR on highly viscous melts.

## 5.3 Temperature Compensation

All sources of magnetic fields are temperature dependent and stable magnetic fields always require temperature control. While for superconducting magnets the boiling temperature of the cryogen depends only on ambient pressure, resistive electro- and permanent-magnets typically need active temperature control or locking systems [66].

**Fig. 14** Coaxial arrangement and rotation of Halbach dipoles (green, left column) and quadrupoles (red, right column). Their magnetization is indicated only by their poles (white or black encircled arrows): **a**, **b** two Halbach di- and quadrupoles which produce the same field strength, $B$, (central green arrow) in case of the quadrupoles the same gradient strength, $G$. Note that only the horizontal component of $B_Q$ is displayed by a red arrow and that this is the derivative of the field (small blue arrows in the center). In the following figures (**c–e**) **a** and **b** are coaxially nested and the outer ring is rotated by an angle $\alpha$. **c** For $\alpha = 0°$ the fields are parallel and the two dipole fields add to $2B$. **d** For $\alpha = 90°$ the fields are orthogonal and the two dipole field vectors add to $\sqrt{2}B$ at an angle of 45°. **e** For $\alpha = 180°$ the fields are antiparallel and cancel each other. The same holds for the gradient in the right column at half the angles, because the gradient rotates at twice the angle of the quadrupole (cf. Eq. (27)). **f** The angular dependence of the combined field magnitude of both dipoles, $B^\Sigma = 2B \cos(\alpha/2)$ (green line) and the two quadrupoles $G^\Sigma = 2G \cos\alpha$ (red line) (color figure online)

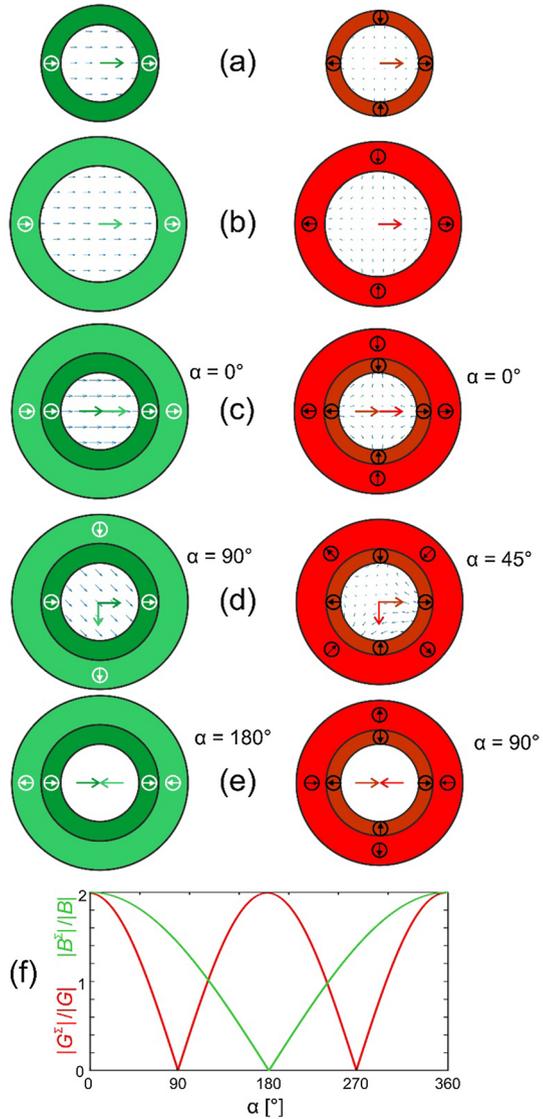

However, Halbach systems can be constructed in a temperature compensated way. This very clever concept was proposed by Danieli et al. [67]. It uses the fact that the two main materials for rare-earth magnets (neodymium and samarium-cobalt) have temperature coefficients, $\kappa$, of their remanence, which are different by a factor 2–4 (see Appendix A). They are defined via

**Fig. 15** Temperature compensated Halbach dipole with the same construction principle as in Fig. 11. However, the system is combined now from two antiparallel dipoles. They are sketched on the right (cyan: NdFeB and yellow: SmCo) next to a measurement of the change of the magnetic field, $\Delta B$, with temperature for the three magnet configurations. The dashed line corresponds to the magnet made only from trapezoidal SmCo blocks, while the dotted line represents the array built from rectangular NdFeB pieces. The continuous line shows the temperature compensated magnet. Adapted from [67] and reproduced with permission of the American Physical Society (color figure online)

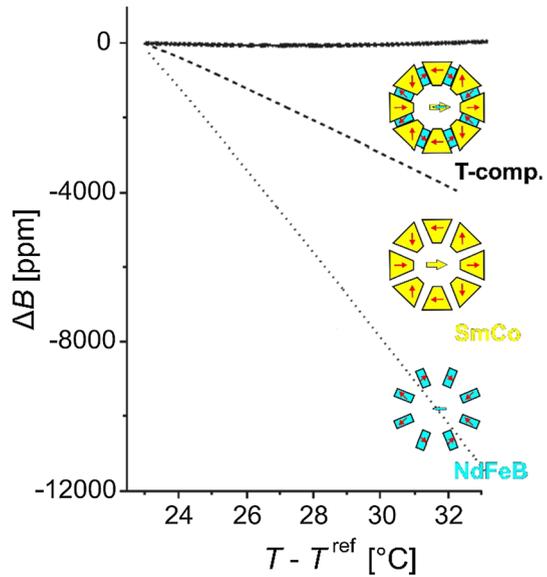

$$B(T) = B^{\text{ref}}(1 + \kappa \, \Delta T) \quad \text{with} \quad \Delta T = T - T^{\text{ref}}, \tag{28}$$

where $B^{\text{ref}}$ is a reference magnetic flux at the temperature $T^{\text{ref}}$. If two Halbach rings, which are made from two materials (A and B) with very different $\kappa$, are combined such that their fields are antiparallel and their strength correspond to the ratios of their $\kappa$, a change in temperature will shift the one field by the same amount as the other, but since they are subtracted the combined field stays constant (cf. Fig. 15).

$$
\begin{aligned}
B^{\Sigma}(T) &= B_{\text{A}}(T) - B_{\text{B}}(T) = B_{\text{A}}^{\text{ref}}\big(1 + \kappa_{\text{A}} \, \Delta T\big) - B_{\text{B}}^{\text{ref}}\big(1 + \kappa_{\text{B}} \, \Delta T\big) \\
&= B_{\text{A}}^{\text{ref}} - B_{\text{B}}^{\text{ref}} + \underbrace{\big(B_{\text{A}}^{\text{ref}}\kappa_{\text{A}} - B_{\text{B}}^{\text{ref}}\kappa_{\text{B}}\big)}_{=0} \, \Delta T, \\
\Rightarrow \quad & \frac{B_{\text{A}}^{\text{ref}}}{B_{\text{B}}^{\text{ref}}} = \frac{\kappa_{\text{B}}}{\kappa_{\text{A}}}.
\end{aligned}
\tag{29}
$$

The original publication also takes thermal expansion into account, which was neglected here for the sake of simplicity.

Temperature stabilization by this approach is a quite general concept, which can be adapted to various magnet geometries. In [67] it was also discussed for quadrupolar Halbach rings and planar Halbach arrays as used in undulators. However, the concept reduces the field by a factor of $(1 - \kappa_{\text{B}}/\kappa_{\text{A}})$. Recently new $Sm_2Co_{17}$ materials have become available which are completely temperature insensitive at RT. Although they have only remanence values in the regime of 0.8–0.9 T, this may still allow to generate higher fields than the approach discussed above [68].

## 5.4 Some Practical Considerations for Construction

The design of a Halbach array should start with an estimation of the necessary field and/or gradient strength needed for the planned experiment. Naturally, the necessary boundary conditions like size, homogeneity, temperature variations, and possible costs have to be determined as well. Then the equations in Sect. 3 should be used to estimate the sizes and the shapes of the magnets. Usually there are conditions that request compromises in field, homogeneity, space, and costs. The necessary equations are easy to program or can be copied from [69]. When a geometry seems to be suitable, the next step is a fine-tuning using some software that considers the actual magnet shapes (and not only their effect on the central field). This can be a simulation package that uses finite-element (e.g. COMSOL Multiphysics) or boundary-element methods (e.g. Amperes, IES, Winnipeg, Canada). If the design includes materials with non-linear magnetization (e.g. iron), the material properties must be properly included in the simulation. Quite often one has to request *BH*-curves from producers. However, there is also very useful freeware available, e.g. FEMM (https://www.femm.info/) which is limited to two dimensions, but includes the use of *BH*-curves. Another great package is magpylib (https://magpylib.readthedocs.io) [70], which is 3D and uses analytical solutions of a set of geometries, hence it is very fast, but needs some elementary skills in Python programming (it also does not include permeabilities and non-linear magnetic behavior). Regardless of which software package is used one should always predetermine the necessary accuracy up to which the geometric optimization is reasonable. Typically, this is limited by the precision to which magnets can be produced and supports machined (normally not better than 100 μm).

After optimizing the geometry and position of the magnets, the support needs some thought, ideally together with strategies on the procedure of mounting the magnets.

Mounting rare-earth magnets of sizes larger than ca. 10 cm$^3$ can already become dangerous, as the forces roughly scale as [4] (for small distances to the surface of two magnets with surface area, *A*)

$$F \approx \frac{B_R A}{2\mu_0}.$$

(30)

Several of the software packages also allow the calculation of the forces, but the direction must be also considered, because a repelled magnet needs some support and fixture to keep it at the position. A good advice is therefore a test run on a scaled down mockup system (e.g. using inexpensive magnets from internet stock and 3D-printed plastic supports). This helps tremendously to figure out if the procedures are feasible before ordering expensive custom-made magnets.

For instance, Eq. (30) predicts a force in the range of 100 N for ca. 1 cm$^2$, while it amounts already to 1000 N for 10 cm$^2$ sized FeNdB magnets. This is just the force between two magnets, two completely assembled rings can easily generate forces that exceed 10 kN. Therefore, upscaling of such magnetic designs quickly becomes an engineering problem in calculating and managing forces, using suitable materials of sufficient thickness and safe procedures for placing and fixing the permanent magnets. This is a process, which requires a lot of experience that can only be acquired safely by starting and tinkering with smaller systems.

For the construction a magnetic set of tools is more useful than expensive nonmagnetic instruments. This is because in a well-planned construction iron plates underneath the support for the permanent magnets can work as a yoke and help to hold the magnets in place while the glue is setting. The same holds for magnetic pliers or clamps which can be placed over the poles of a magnet to shield it from the others during positioning, thus significantly reducing mutual forces [34]. Nevertheless, a good magnet manufacturer also has a larger supply of wooden blocks, plates and wedges (and some band-aid). The handiest tool of all though is a 3D-compass or pole-finder to quickly check the polarity of a magnet or entire arrangements.

Since magnetic resonance does not only require magnets, readers, who are interested in developing or building equipment (amplifiers, coils, magnets, software, etc.) are referred to the webpage [71] of the Open Source Imaging (OSI[2]) initiative [72].

## 6 Conclusion

The intention of this review on Halbach magnets for applications in magnetic resonance was to compile the necessary knowledge to plan, design and construct such arrays. Special emphasis was given to simple, analytical estimations of magnetic field strength or their gradients. Another focus was to summarize various strategies to homogenize the fields as this is of paramount importance for NMR. As to the topic of homogeneity, to provide a sufficiently homogeneous field in a given space is a difficult task, especially with permanent magnets, because their physical properties usually vary in the percent range, as has been shown here. Thus, without further measures, it would be a matter of coincidence if a homogeneity requirement is met, based only on an even perfect numerical design of the magnet arrangement. Several such measures have been mentioned in this publication, they comprise the appropriate placement of the permanent magnets in the arrangement such that their magnetic property variations cancel, the adaptive placement of shim magnets or steel disks/ sheets on the poles and even the active shimming by adjustable coil currents. Homogeneity is difficult to achieve if space restrictions force the magnetic arrangement to be small, the homogeneous region scales with the size of the arrangement, but the size of the permanent magnets and the forces between them scale more than linearly, which may become a safety issue. In the end, it may only be possible to achieve a required homogeneity with a combination of all these measures. A good numerical design, as described in many publications, can only be the starting point for further experimental optimization.

The last section is a collection of features that are to some extent special for Halbach magnets and allows to use them in unique experiments. We hope that this compilation enables interested experimentalists to create dedicated magnets for their special needs, and maybe share a bit of the fascination for these 'magic rings'.

## Appendix A: Typical Properties of Rare-Earth Magnet Materials

See Table 1.

**Table 1** Typical values for neodymium and samarium-cobalt magnetic materials: the temperature coefficients, $\kappa$, for the remanence and coercivity are typical values in the temperature range around room temperature (ca. $0-80$ °C)

| Material grade | $Nd_2Fe_{14}B$ | | | $SmCo_5$ | $Sm_2Co_{17}$ |
|---|---|---|---|---|---|
| | N45 | N48 | N54 | | |
| Remanence, $B_R$ [T] | 1.32 to 1.38 | 1.38 to 1.42 | 1.45 to 1.50 | 0.95 to 1.0 | 1.08 to 1.15 |
| Coercivity, $H_c$ [kA/m] | 2000 | 1600 | 880 | 1200 to 1800 | 500 to 720 |
| Temp. coeff. $B_R$, $\kappa$ [ppm/K] | $-1200$ | $-1200$ | $-1200$ | $-350$ to $-500$ | $-300$ to $-450$ |
| Temp. coeff. $H_c$, $\kappa$ [ppm/K] | $-4650$ | $-6000$ | $-7500$ | $-1500$ to $-3000$ | $-2000$ to $-3000$ |
| Rel. permeability, $\mu_r$ | 1.05 | 1.05 | 1.05 | 1.05 | 1.05 to 1.1 |
| Density [g/cm$^3$] | 7.5 | 7.6 | 7.6 | 8.1 to 8.4 | 8.3 to 8.5 |
| Max. temp. of use [°C] | 180 | 120 | 80 | 250 | 350 |

The maximal temperature of use is not the Curie temperature, which is above 300 °C for neodymium and 700 °C for samarium-cobalt magnets. Please be aware that particularly for SmCo magnets the range of properties is very large. The values given here are taken from the fraction with high remanences

## Appendix B: Mandhala Geometry

(a)  $N$ and $R_c$ given:

The geometry is illustrated in Fig. 16. When the number of magnets, $N$, and the central radius, $R_c$ are given, Table 2 sums up the construction parameters. A distance between neighboring magnets, $g$, is also introduced, maximally dense packing is then for $g = 0$ [35].

Then the inner and outer radius of the arrangement are given by

$$R_i = R_c - \frac{d + g}{2} \quad \text{and} \quad R_o = R_c + \frac{d + g}{2}.$$

Size, side length and area are listed in Table 2.

**Fig. 16** Illustration of the geometric symbols in Table 2 (color figure online)

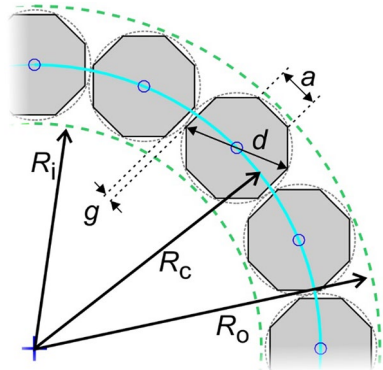

**Table 2** Dense packing of Mandhalas from magnets with circular, square and regular $N/2$-polygonal (i.e. regular polygons with half the number of vertices as magnets in the ring) footprint. Listed are the magnet size, $d$, (diameter for circles, diagonal for squares, long diagonal for polygons), side length $a$, and their area, $A_M$. See also Fig. 16. The last column is for magnets with footprints of regular polygons with $K$ vertices. They are not necessarily optimally dense packed

|  | Circles | Squares | $N/2$-polygons | $K$-polygon |
|---|---|---|---|---|
| $d$ | $2R_c \sin \frac{\pi}{N} - g$ | $2R_c\, \Psi(N) - g$ | $2R_c\, \Theta(N) - g$ | $2R_c \sin \frac{\pi}{N} - g$ |
| $a$ | — | $\frac{d}{\sqrt{2}}$ | $d \sin \frac{2\pi}{N}$ | $d \sin \frac{\pi}{K}$ |
| $A_M$ | $\frac{N\pi}{4} d^2$ | $\frac{N}{2} d^2$ | $\frac{N^2}{16} d^2 \sin \frac{4\pi}{N}$ | $\frac{NK}{8} d^2 \sin \frac{2\pi}{K}$ |

with 
$$\Psi(N) \equiv \frac{\cos \frac{2\pi}{N} - \sin \frac{2\pi}{N} - \sqrt{2}\sin(\frac{\pi}{4} - \frac{4\pi}{N})}{\sqrt{2}\cos(\frac{\pi}{4} - \frac{4\pi}{N}) + 1}$$
and 
$$\Theta(N) \equiv \begin{cases} \frac{1}{2}\tan \frac{2\pi}{N} & \text{for } \frac{N}{2} \text{ even} \\ 2\frac{\sin \frac{\pi}{N}}{\cos \frac{2\pi}{N} + 1} & \text{for } \frac{N}{2} \text{ odd} \end{cases}$$

(b) $R_i$ and $R_o$ given:

This is a less determined set, but very useful for fitting a magnet system between inner and outer limits.

From Eq. (12) $R_c = \frac{R_i + R_o}{2}$, and for circular footprints this gives a diameter $d = R_o - R_i - g/2$, while for polygons with $K$ vertices the long diagonal is given by $d = (2R_o - 2R_i - g) \sin(\pi/K)$.

Note that for an ill-chosen set, overlapping structures are possible. The overlapping must then by avoided by adjusting $N = \left\lceil \pi / \sin^{-1}\left(\frac{d+g}{R_o + R_i}\right) \right\rceil \in \mathbb{N}^2$ or $g$.

---

[2] This notation means that it is rounded to the next smaller integer value.

# Appendix C: Reduction Factor $f^L$ for Various Values of $k$

See Table 3.

**Table 3** Calculated reduction factor $f^L(k)$, cf. Eq. (14) in Sect. 3.4 due to the truncation of Halbach multipoles to length $L$, with $\Phi = L^2 + 4R_c^2$. The explicit solutions of the general solution in Eq. (31)

| $k$ | $f^L(k)$ |
|---|---|
| 1 | $L(L^2 + 6R_c^2)\,\Phi^{-3/2}$ |
| 2 | $L(L^4 + 10L^2R_c^2 + 30R_c^4)\,\Phi^{-5/2}$ |
| 3 | $L(L^6 + 14L^4R_c^2 + 70L^2R_c^4 + 140R_c^6)\,\Phi^{-7/2}$ |
| 4 | $L(L^8 + 18L^6R_c^2 + 126L^4R_c^4 + 420L^2R_c^6 + 630R_c^8)\,\Phi^{-9/2}$ |
| 5 | $L(L^{10} + 22L^8R_c^2 + 198L^6R_c^4 + 924L^4R_c^6 + 2310L^2R_c^8 + 2772R_c^{10})\,\Phi^{-11/2}$ |
| 6 | $L(L^{12} + 26L^{10}R_c^2 + 286L^8R_c^4 + 1716L^6R_c^6 + 6006L^4R_c^8 + 12012L^2R_c^{10} + 12012R_c^{12})\,\Phi^{-13/2}$ |

The general solution is
$$f^L(k) = \frac{L}{\sqrt{L^2 + 4R_c^2}} \sum_{j=0}^{k} \binom{2j}{j} \left( \frac{L^2}{R_c^2} + 4 \right)^{-j}, \tag{31}$$

with the binomial coefficient $\binom{2j}{j} = \dfrac{(2j)!}{(j!)^2}$.

# Appendix D: Comparison of Analytical Solution to FEM Simulation

See Table 4.

**Table 4** Comparison of the analytically obtained estimations of the magnetic flux density in the center of a Halbach dipole and quadrupole ('theory' from Sect. 3) to values obtained from a FEM simulation (COMSOL Multiphysics 5.5) for $R_i = 10$ mm, $R_o = 15$ mm, $B_R = 1$ T

| | Dipole ($k=1$) | | Quadrupole ($k=2$) | |
|---|---|---|---|---|
| | $B_{theory}$ [mT] | $B_{FEM}$ [mT] | $G_{theory}$ [T/m] | $G_{FEM}$ [T/m] |
| Ideal (Eq. (6)) | 405.465 | 405.461 | 66.667 | 66.664 |
| $\mu_r = 1.1$ (Eq. (8)) | 386.547 | 386.215 | 63.556 | 63.517 |
| Segmented $N=16$ (Eq. (10)) | 376.688 | 376.368 | 59.944 | 59.908 |
| Octagons $d=1.98$ mm (Eq. (11)) | 290.935 | 294.308 | 46.298 | 47.103 |
| Truncated to $L=5$ mm (Eq. (14)) | 84.488 | 88.745 | 16.593 | 17.482 |

The rows of the table follow the sequence of the main text in Sect. 3, starting with the ideal 2D Halbach ring, introducing permeability, then segmenting it, changing the shape of segments from cylindrical to octagons, and finally truncating it in the third dimension

## Appendix E: Optimal Arrangement of Individual Magnets in a Halbach Ring

We start from a set of $K$ magnets each with an individual magnetic moment, $\boldsymbol{m}_j$, and an average of $\overline{\boldsymbol{m}} = 1/K \sum_{j=1}^{K} \boldsymbol{m}_j$ (cf. Fig. 7b). The individual deviation of each magnet from the mean is then a deviation vector $\Delta\boldsymbol{m}_j = \left(\overline{m}_x - m_{xj}, \overline{m}_y - m_{yj}, \overline{m}_z - m_{zj}\right) = \left(\Delta m_{xj}, \Delta m_{yj}, \Delta m_{zj}\right)$ to include angular misalignments (however, the polarization directions of the magnets remain in the $xy$-plane). If an arbitrarily chosen ensemble of $N$ magnets is then arranged to a Halbach multipole each magnet and hence $\Delta\boldsymbol{m}_j$ must be rotated according to Eq. (4) by the following rotation matrix, $\mathbf{R}$

$$\mathbf{R} = \begin{pmatrix} \cos\left((k+1)\theta_j\right) & -\sin\left((k+1)\theta_j\right) & 0 \\ \sin\left((k+1)\theta_j\right) & \cos\left((k+1)\theta_j\right) & 0 \\ 0 & 0 & 1 \end{pmatrix} \quad \text{with } \theta_j = \frac{2\pi}{N}j. \quad (32)$$

$$\delta\boldsymbol{m}_j = \mathbf{R}\,\Delta\boldsymbol{m}_j = \begin{pmatrix} \cos\left((k+1)\theta_j\right)\Delta m_{xj} - \sin\left((k+1)\theta_j\right)\Delta m_{yj} \\ \sin\left((k+1)\theta_j\right)\Delta m_{xj} + \cos\left((k+1)\theta_j\right)\Delta m_{yj} \\ \Delta m_{zj} \end{pmatrix} \quad \text{and } \boldsymbol{r}_j = \begin{pmatrix} R_c\cos\theta_j \\ R_c\sin\theta_j \\ z \end{pmatrix}, \quad (33)$$

where $\boldsymbol{r}_j$ is the position vector and the magnet is situated at a position $z$ in axial direction. When treated as a magnetic dipole it produces a "deviation" field in the center given by (cf. Eq. (23))

$$\delta\boldsymbol{B}_j(0,0,z) = \frac{\mu_0}{4\pi}\frac{3\boldsymbol{r}_j(\delta\boldsymbol{m}_j \cdot \boldsymbol{r}_j) - \delta\boldsymbol{m}_j r_j^2}{r_j^5}$$

$$= C \begin{pmatrix} 2[-\cos(\hat{k}\theta_j)\Delta m_{xj} + \sin(\hat{k}\theta_j)\Delta m_{yj}]z^2 + 6R_c\cos\theta_j\,\Delta m_{zj}z + \cdots \\ -2[\sin(\hat{k}\theta_j)\Delta m_{xj} + \cos(\hat{k}\theta_j)\Delta m_{yj}]z^2 + 6R_c\sin\theta_j\,\Delta m_{zj}z + \cdots \\ 4\Delta m_{zj}z^2 + 6R_c[\cos(k\theta_j)\Delta m_{xj} - \sin(k\theta_j)\Delta m_{yj}]z - \cdots \\ \cdots + R_c^2\{[\cos(\hat{k}\theta_j) + 3\cos(\check{k}\theta_j)]\Delta m_{xj} - [\sin(\hat{k}\theta_j) + 3\sin(\check{k}\theta_j)]\Delta m_{yj}\} \\ \cdots + R_c^2\{[\sin(\hat{k}\theta_j) - 3\sin(\check{k}\theta_j)]\Delta m_{xj} + [\cos(\hat{k}\theta_j) - 3\cos(\check{k}\theta_j)]\Delta m_{yj}\} \\ \cdots - 2R_c^2\Delta m_{zj} \end{pmatrix} \quad (34)$$

$$\text{with} \quad C \equiv \frac{\mu_0}{8\pi(R_c^2 + z^2)^{5/2}}, \quad \hat{k} \equiv k+1 \quad \text{and} \quad \check{k} \equiv k-1.$$

In a ring made from $N$ ($\leq K$) magnets an ideal arrangement is then found if the erroneous contributions to the magnetic field due to the deviations $\Delta\boldsymbol{m}_j$ cancel for each direction and each power of $z$. Hence, one gets 8 sums, which must be minimized numerically (symbolized by min $\{\dots\}$).

$$
\begin{aligned}
\text{(I)} \quad & \min\left\{ \left| \sum_{j=1}^{N} \left[ -\cos\left((k+1)\theta_j\right)\Delta m_{xj} + \sin\left((k+1)\theta_j\right)\Delta m_{yj} \right] \right| \right\} \ \wedge \\
\text{(II)} \quad & \min\left\{ \left| \sum_{j=1}^{N} \cos\theta_j \Delta m_{zj} \right| \right\} \ \wedge \\
\text{(III)} \quad & \min\left\{ \left| \sum_{j=1}^{N} \left[ \left(\cos\left((k+1)\theta_j\right) + 3\cos\left((k-1)\theta_j\right)\right)\Delta m_{xj} - \left(\sin\left((k+1)\theta_j\right) + 3\sin\left((k-1)\theta_j\right)\right)\Delta m_{yj} \right] \right| \right\} \ \wedge \\
\text{(IV)} \quad & \min\left\{ \left| \sum_{j=1}^{N} \left[ \sin\left((k+1)\theta_j\right)\Delta m_{xj} + \cos\left((k+1)\theta_j\right)\Delta m_{yj} \right] \right| \right\} \ \wedge \\
\text{(V)} \quad & \min\left\{ \left| \sum_{j=1}^{N} \sin\theta_j \Delta m_{zj} \right| \right\} \ \wedge \\
\text{(VI)} \quad & \min\left\{ \left| \sum_{j=1}^{N} \left[ \left(\sin\left((k+1)\theta_j\right) - 3\sin\left((k-1)\theta_j\right)\right)\Delta m_{xj} + \left(\cos\left((k+1)\theta_j\right) - 3\cos\left((k-1)\theta_j\right)\right)\Delta m_{yj} \right] \right| \right\} \ \wedge \\
\text{(VII)} \quad & \min\left\{ \left| \sum_{j=1}^{N} \Delta m_{zj} \right| \right\} \ \wedge \\
\text{(VIII)} \quad & \min\left\{ \left| \sum_{j=1}^{N} \left[ \cos k\theta_j \ \Delta m_{xj} - \sin k\theta_j \ \Delta m_{yj} \right] \right| \right\}.
\end{aligned}
\tag{35}
$$

The best arrangement of magnets is then the one for which all these sums are minimal. There are different strategies to compute this, e.g. using a Monte Carlo approach [73] or arrange $N-1$ magnets and then look for a close to perfect match among the rest of magnets.

In the calculations above a magnetic moment vector was used to characterize the magnets or their deviations, respectively. Practically, this can be achieved by determining a value that scales with the magnetic moment, e.g., the magnetic flux at an arbitrarily chosen but fixed distance sampled by a 3D-Hall probe as suggested in Fig. 7a. However, this experiment has to be very reproducible. Great care has to be taken to avoid geometric misalignments or temperature drifts (even from keeping a magnet in the hands for too long). Such values can then be used to optimize one or several rings by minimizing Eq. (35). If a set of rings should be stacked like in Sect. 4.2, it must be checked if they have slightly different central field strengths. To optimize homogeneity in the center of a stack of these rings one strategy is to place the rings with highest homogeneity in the $xy$-plane in the center of that stack and optimize their position along $z$ according to the concept discussed in Sect. 4.2. For instance, searching for a numerical minimum of a sum of equations of the type of Eq. (20).

## Appendix F: Optimized Stack Distances

Table 5 gives the optimized distances $s_j$ of stacked Halbach multipoles for $k = 1$–6. All optimized to an accuracy better than $10^{-3}$ and relative to $R_c$ (cf. [37] and Fig. 8). For instance, the optimal distance between two ring centers is then $2s_1 R_c = 0.824\,R_c$ for $k = 1$. The last but one column gives the increase of the field (or its $(k-1)^{th}$ derivative) by the factor, $f^{zs}$, in the center of the stack relative to that of a single ring. The last column gives the central region, $\Delta z^{hom}$, on the axis where the field (or its $(k-1)^{th}$ derivative) does not vary more than $10^{-3}$ from the central value.

**Table 5** Optimized distances $s_j$ of Halbach multipoles ($k$ ranging from 1 to 6) stacked from $n=2$ to 12 rings

| $n$ | $s_1/R_c$ | $s_2/R_c$ | $s_3/R_c$ | $s_4/R_c$ | $s_5/R_c$ | $s_6/R_c$ | $f^{zs}$ | $\Delta z^{hom}/R_c$ |
|---|---|---|---|---|---|---|---|---|
| $k=1$ | | | | | | | | |
| 2 | $\pm 0.4120$ | – | – | – | – | – | 1.351 | 0.3214 |
| 4 | $\pm 0.3027$ | $\pm 0.7764$ | – | – | – | – | 2.221 | 0.4011 |
| 6 | $\pm 0.2528$ | $\pm 0.7983$ | $\pm 1.1322$ | – | – | – | 2.550 | 1.2556 |
| 8 | $\pm 0.2244$ | $\pm 0.6691$ | $\pm 1.1526$ | $\pm 1.3697$ | – | – | 2.946 | 1.6236 |
| 10 | $\pm 0.2225$ | $\pm 0.6664$ | $\pm 1.1098$ | $\pm 1.5691$ | $\pm 1.8185$ | – | 2.981 | 2.3679 |
| 12 | $\pm 0.2230$ | $\pm 0.6701$ | $\pm 1.1150$ | $\pm 1.5584$ | $\pm 2.0282$ | $\pm 2.2587$ | 2.975 | 3.3172 |
| $k=2$ | | | | | | | | |
| 2 | $\pm 0.3589$ | – | – | – | – | – | 1.309 | 0.3102 |
| 4 | $\pm 0.2921$ | $\pm 0.4534$ | – | – | – | – | 2.541 | 0.3218 |
| 6 | $\pm 0.2392$ | $\pm 0.7336$ | $\pm 1.1303$ | – | – | – | 2.201 | 1.3063 |
| 8 | $\pm 0.2240$ | $\pm 0.6674$ | $\pm 1.1282$ | $\pm 1.4693$ | – | – | 2.385 | 1.9791 |
| 10 | $\pm 0.2209$ | $\pm 0.6669$ | $\pm 1.1058$ | $\pm 1.5684$ | $\pm 1.8966$ | – | 2.402 | 2.8620 |
| 12 | $\pm 0.2089$ | $\pm 0.6262$ | $\pm 1.0442$ | $\pm 1.4589$ | $\pm 1.8912$ | $\pm 2.1929$ | 2.553 | 3.3292 |
| $k=3$ | | | | | | | | |
| 2 | $\pm 0.3208$ | – | – | – | – | – | 1.287 | 0.2776 |
| 4 | $\pm 0.2712$ | $\pm 0.3861$ | – | – | – | – | 2.523 | 0.2919 |
| 6 | $\pm 0.2179$ | $\pm 0.6645$ | $\pm 1.0478$ | – | – | – | 2.080 | 1.2082 |
| 8 | $\pm 0.2086$ | $\pm 0.6236$ | $\pm 1.0474$ | $\pm 1.4024$ | – | – | 2.194 | 1.8849 |
| 10 | $\pm 0.2064$ | $\pm 0.6219$ | $\pm 1.0334$ | $\pm 1.4582$ | $\pm 1.8033$ | – | 2.208 | 2.7260 |
| 12 | $\pm 0.2004$ | $\pm 0.5993$ | $\pm 1.0003$ | $\pm 1.3978$ | $\pm 1.8074$ | $\pm 2.1382$ | 2.285 | 3.3176 |
| $k=4$ | | | | | | | | |
| 2 | $\pm 0.2927$ | – | – | – | – | – | 1.273 | 0.2535 |
| 4 | $\pm 0.2545$ | $\pm 0.3410$ | – | – | – | – | 2.508 | 0.2687 |
| 6 | $\pm 0.2024$ | $\pm 0.6148$ | $\pm 0.9831$ | – | – | – | 1.995 | 1.1361 |
| 8 | $\pm 0.1903$ | $\pm 0.5696$ | $\pm 0.9545$ | $\pm 1.2919$ | – | – | 2.137 | 1.6834 |
| 10 | $\pm 0.1948$ | $\pm 0.5860$ | $\pm 0.9748$ | $\pm 1.3718$ | $\pm 1.7168$ | – | 2.082 | 2.6041 |
| 12 | $\pm 0.1944$ | $\pm 0.5828$ | $\pm 0.9715$ | $\pm 1.3595$ | $\pm 1.7514$ | $\pm 2.1025$ | 2.090 | 3.2618 |
| $k=5$ | | | | | | | | |
| 2 | $\pm 0.2708$ | – | – | – | – | – | 1.263 | 0.2333 |
| 4 | $\pm 0.2159$ | $\pm 0.3470$ | – | – | – | – | 2.443 | 0.2403 |
| 6 | $\pm 0.1892$ | $\pm 0.5736$ | $\pm 0.9244$ | – | – | – | 1.942 | 1.0706 |
| 8 | $\pm 0.1589$ | $\pm 0.4762$ | $\pm 0.7972$ | $\pm 1.0797$ | – | – | 2.325 | 1.1387 |
| 10 | $\pm 0.1670$ | $\pm 0.5027$ | $\pm 0.8358$ | $\pm 1.1777$ | $\pm 1.4694$ | – | 2.207 | 2.1049 |
| 12 | $\pm 0.1700$ | $\pm 0.5101$ | $\pm 0.8502$ | $\pm 1.1899$ | $\pm 1.5328$ | $\pm 1.8417$ | 2.172 | 2.7116 |
| $k=6$ | | | | | | | | |
| 2 | $\pm 0.2533$ | – | – | – | – | – | 1.255 | 0.2194 |
| 4 | $\pm 0.2422$ | $\pm 0.2684$ | – | – | – | – | 2.491 | 0.2461 |
| 6 | $\pm 0.1782$ | $\pm 0.5397$ | $\pm 0.8745$ | – | – | – | 1.904 | 1.0029 |
| 8 | $\pm 0.1736$ | $\pm 0.5202$ | $\pm 0.8692$ | $\pm 1.1950$ | – | – | 1.965 | 1.5626 |
| 10 | $\pm 0.1733$ | $\pm 0.5203$ | $\pm 0.8668$ | $\pm 1.2155$ | $\pm 1.5415$ | – | 1.966 | 2.2333 |

**Table 5** (continued)

| $n$ | $s_1/R_c$ | $s_2/R_c$ | $s_3/R_c$ | $s_4/R_c$ | $s_5/R_c$ | $s_6/R_c$ | $f^{zs}$ | $\Delta z^{hom}/R_c$ |
|---|---|---|---|---|---|---|---|---|
| 12 | ±0.1684 | ±0.5052 | ±0.8421 | ±1.1788 | ±1.5169 | ±1.8348 | 2.024 | 2.7019 |

## Appendix G: Optimal Opening Angles for Mandhalas

The optimum opening angle for a Mandhala ring can be calculated in the dipole approximation, in which the magnetic field of every magnet is reduced to that of a dipole and only nearest neighbor interactions between dipoles are considered. This assumption is justified by the fact that the distance to next-but-one neighbors are already double the distance to the nearest neighbors and the forces between the magnets decrease inversely with the fourth power of the distance, as shown below. Thus, nearest neighbor interactions are at least 16 times stronger compared to that of any other pair in the ring.

The force exerted on one dipole with dipole moment $\boldsymbol{m}_1$ by another dipole characterized by $\boldsymbol{m}_2$ is given by

$$\boldsymbol{F} = \nabla\left(\boldsymbol{m}_1 \frac{\mu_0}{4\pi} \frac{3(\boldsymbol{m}_2 \cdot \boldsymbol{d})\boldsymbol{d} - d^2\boldsymbol{m}_2}{d^5}\right) = \frac{\mu_0}{4\pi}\nabla\left(\frac{3(\boldsymbol{m}_1 \cdot \boldsymbol{d})(\boldsymbol{m}_2 \cdot \boldsymbol{d})}{d^5} - \frac{(\boldsymbol{m}_1 \cdot \boldsymbol{m}_2)}{d^3}\right), \tag{36}$$

where $\boldsymbol{d}$ denotes the vector connecting the two dipoles and $d = |\boldsymbol{d}|$ is their distance, which will become the diameter of a corresponding sphere or cylinder later on (see Fig. 17a, b). The application of the gradient operator yields the following expression for the force

$$\boldsymbol{F} = \frac{3\mu_0 m_1 m_2}{2\pi d^4} \frac{(\boldsymbol{e}_{m_1} \cdot \boldsymbol{e}_d)\boldsymbol{e}_{m_2} + (\boldsymbol{e}_{m_2} \cdot \boldsymbol{e}_d)\boldsymbol{e}_{m_1} - \left(5(\boldsymbol{e}_{m_1} \cdot \boldsymbol{e}_d)(\boldsymbol{e}_{m_2} \cdot \boldsymbol{e}_d) - (\boldsymbol{e}_{m_1} \cdot \boldsymbol{e}_{m_2})\right)\boldsymbol{e}_d}{2}, \tag{37}$$

where $m_1$ and $m_2$ are the absolute values of the magnetic moments and $\boldsymbol{e}_d, \boldsymbol{e}_{m_1}, \boldsymbol{e}_{m_2}$ are the unit vectors of the vectors specified in their subscripts. In this representation the amplitude of the force and its angular dependence are clearly separated. Also, the force between the dipoles has components which may be perpendicular to the vector $\boldsymbol{d}$ connecting them, see the first terms in the nominator.

When arranged in a Mandhala ring, two such dipoles on opposite sides of the opening line would generate a torque, $\tau$, which is calculated based on the component of the force along the direction given by $\boldsymbol{e}_d$. The length of the lever arm is given by $2R_c$

$$\tau = 2R_c \boldsymbol{F} \cdot \boldsymbol{e}_d = 2R_c \frac{3\mu_0 m_1 m_2}{2\pi d^4} \underbrace{\frac{(\boldsymbol{e}_{m_1} \cdot \boldsymbol{e}_{m_2}) - 3(\boldsymbol{e}_{m_1} \cdot \boldsymbol{e}_d)(\boldsymbol{e}_{m_2} \cdot \boldsymbol{e}_d)}{2}}_{:= \Xi}. \tag{38}$$

The last fraction, describing the angular dependence, $\Xi$, is normalized by a factor of 2 in order to restrict its values between $-1$ and $+1$. The coefficients can then be regarded as an amplitude of the torque.

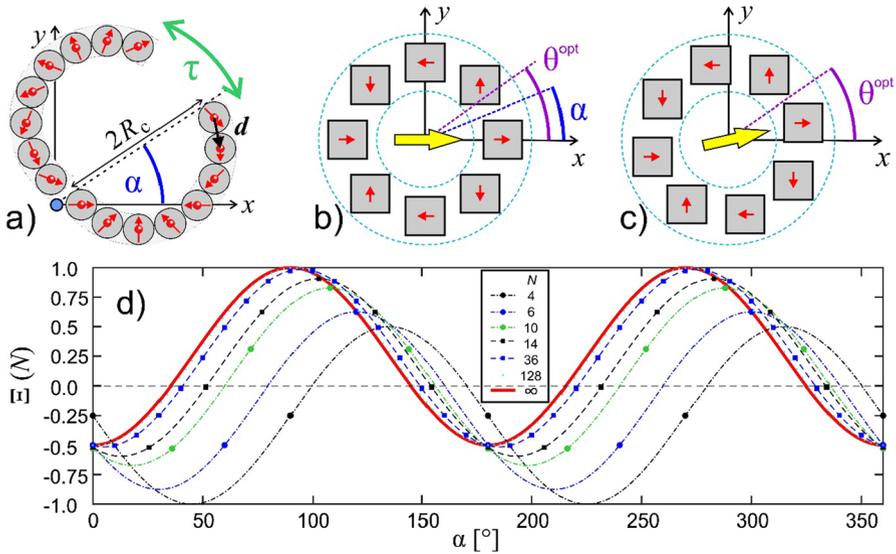

**Fig. 17** **a** Sketch of opening a Mandhala of $N=16$ cylinders at a possible (not the optimal) angle $\alpha$ between two magnets, using a hinge (blue). The green arrow indicates the calculated torque, $\tau$. The red spheres in the centers of each cylinder should remind that the calculation was done for dipoles at a distance $d$ to each other. **b** The difference between the optimal angle, $\theta^{opt}$ from Eq. (25), for force-free opening and $\alpha$ as the angle of a possible opening angle close to $\theta^{opt}$. Shown for $k=1$ and $N=8$ where this difference is pronounced. **c** The same situation but now the magnets are arranged with an additional phase $\varphi=\theta^{opt}-\alpha$ such that $\alpha=\theta^{opt}$ but now the magnetic field (yellow arrow) is tilted by $\varphi$. **d** Illustration of the angular term, $\Xi(1,j,N)$, of the torque (Eq. (41)) for various values of $N$. The markers indicate the possible opening angles $\alpha_j$ between magnets in the ring ($j=0,1\ldots,N$). All the curves are continuous expressions ($j\rightarrow\infty$ in Eq. (41)) for different $N$ to guide the eye. The thick red line shows the limit for $N\rightarrow\infty$. Negative values of $\Xi$ correspond to a force that opens (repelling force) the two halves of the Mandhala, positive values bring them together (attracting force) (color figure online)

If this formula is to be applied to real magnets, their magnetic moments should be related to their respective volumes $V$ via

$$\boldsymbol{m} = \boldsymbol{M}V, \tag{39}$$

where the absolute value of the magnetization $\boldsymbol{M}$ is given by $M = B_R/\mu_0$.

To put the calculated torques into perspective, the maximum torque $\tau$ in a Mandhala ring can be estimated by using the amplitude factor in Eq. (38) and applying some simplifications in order to translate from dipoles to real magnets. The calculated torque values would be best reproduced, if the magnets had the shape of spheres, because homogeneously magnetized spheres generate an exact dipolar field on their outside. For the usual case of cylindrical magnets, the results will be transferable if their height $L$ is chosen such that they have the same volume as their spherical counterparts with the same cross-sectional area, which is fulfilled for the height $L=2/3\ d$. For higher magnets the torque, $\tau$, must be scaled in these units, thus we find for the torque amplitude, $\tau_{cyl}$, for cylindrical magnets

$$\tau_{cyl} \;=\; \frac{L}{\frac{2}{3}d}\,\frac{3\mu_0 m_1 m_2}{2\pi d^4}\,2R_c \;\;\overset{*}{\rightarrow}\;\; \frac{\pi \sin(\frac{\pi}{N})}{4\mu_0}\,B_R^2 L R_c^2$$

$$* \text{using: } m_{1,2} \;=\; \frac{B_R V}{\mu_0}, \quad V \;=\; \frac{\pi d^3}{6}, \quad \text{and } d \;\overset{\text{Tab. 2}}{=}\; 2R_c \sin\frac{\pi}{N}. \tag{40}$$

The angular dependence, $\Xi$, of the torque given in the underbrace in Eq. (38) can be evaluated for two neighboring magnets in the ring with indices $j$ and $j+1$ (see Eq. (32)), respectively, in order to identify the set of magnet numbers $N$ in a ring which lend themselves best for force-free opening. This gives

$$\Xi(k, j, N) \;=\; \frac{1}{4}\left[\cos\left(\frac{2\pi(k+1)}{N}\right) - 3\cos\left(\frac{2\pi k}{N}(1+2j)\right)\right]. \tag{41}$$

Figure 17d shows this dependence for several values of $N$. The maximum contribution of this angular expression to the torque is a factor of 1. Obviously, angles with close to vanishing torque can be realized (e. g. for $N=14$). Higher values of $N$ may also give interesting configurations, but it must be kept in mind that an increase in $N$ means a reduction of the magnet size and a reduced field in the center. To give an example for the order of magnitude of a typical range of torque in a Mandhala dipole made from $N=16$ cylindrical magnets with $B_R=1.4$ T arranged on a circle with $R_c=5$ cm ($d=1.95$ cm, $L=1.3$ cm) Eq. (40) yields a torque amplitude of $\tau_{cyl}=7.77$ Nm. The angular dependence yields the torque closest to zero $\tau_{min}=-0.86$ Nm $[-1.13$ Nm$]$ (at $j=1, 6, 9, 14$) and a maximal value of $\tau_{max}=+6.76$ Nm $[+8.47$ Nm$]$ (at $j=3, 4, 11, 12$). For comparison, the values obtained by a numerical simulation are shown in square brackets.

The magnetization directions of two consecutive magnets (dipoles) in a Mandhala ring with many magnets differ very little. If their common angle with their connecting line is denoted by $\theta$, which is also their common angle with respect to the origin, see Fig. 13 a, the angular dependence of the torque according to Eq. (41) becomes

$$\Xi(k, \theta) \;=\; \frac{1}{4}\,(1 - 3\cos(2k\theta)) \;=\; \frac{1}{2}\big(3\sin^2(k\theta) \;-\; 1\big). \tag{42}$$

This expression vanishes for the values, $\theta^{opt}$, given in Eq. (25), where the Mandhala ring can be opened without force in the limit of many dipoles forming the ring.

High torque values are not a real issue for small magnets but can become considerable for larger systems. Therefore, it is easier to tilt the magnetization pattern or the magnet orientation (given by Eq. (4)) by a phase $\varphi = \theta^{opt} - \alpha$, as explained in Fig. 17b, c. This rotates the flux pattern by $\varphi$ but causes the geometrically determined angle $\alpha = \theta^{opt}$ to allow opening the structure without force. A specific phase angle is $\varphi = \frac{1}{2}\frac{2\pi}{N}$, which puts the magnets in the intermediate positions and leads to another symmetric, "conjugate" arrangement. In any case, a nonzero phase angle $\varphi$, will lead to different positions of the markers in Fig. 17d.

**Acknowledgements** PB wants to thanks his former employers, H. W. Spiess (MPI for Polymer Research, Mainz), U. Schurr (Forschungszentrum Jülich), and W. Heil (University of Mainz) for their continuous interest and support in (t)his side activity, which led to some of the presented work. Furthermore, he wants to thank Mike Mallett, Hanspeter Raich, Dagmar van Dusschoten, and Carel Windt for their stimulating ideas, help and friendship. Both authors want to thank Bernhard Blümich for his mentor- and friendship and in particular his continuous stimulation and promotion of the entire field of compact and mobile NMR. This review is dedicated to him on the occasion of his 70th birthday. Finally, we are indebted to the two unknown referees of this manuscript, whose suggestions greatly improved the manuscript in particular Appendix G.

**Author Contributions** PB is responsible for the concept of this review and all figures. Both authors contributed about equally to writing the manuscript and the presented calculations. Both read and approved the final version of the manuscript.

**Funding** Open Access funding enabled and organized by Projekt DEAL. OpenAccess funding enabled and organized by Projekt DEAL.

**Availability of Data and Materials** Not applicable.

## Declarations

**Conflict of interest** All authors declare that they have no conflict of interest.

**Ethical approval** Not applicable.

**Publisher's Note** Springer Nature remains neutral with regard to jurisdictional claims in published maps and institutional affiliations.